\documentclass[modern]{aastex61}
\usepackage{graphicx}
\usepackage{psfig}
\usepackage{natbib}

\newcommand{\gsim}{\mbox{\hspace{.2em}\raisebox{.5ex}{$>$}\hspace{-.8em}\raisebox{-.5ex}{$\sim$}\hspace{.2em}}}
\newcommand{\lsim}{\mbox{\hspace{.2em}\raisebox{.5ex}{$<$}\hspace{-.8em}\raisebox{-.5ex}{$\sim$}\hspace{.2em}}}

\newcommand{\E}[1]{\times 10^{#1}}
\newcommand{\twCO}{$^{12}$CO}  \newcommand{\thCO}{$^{13}$CO}

\newcommand{\HII}{\mbox{H\,\textsc{ii}}}

      \newcommand{\ps}{\,{\rm s}^{-1}}
    \newcommand{\Msun}{M_{\odot}}   
    \newcommand{\km}{\,{\rm km}}

\begin{document}

\title{
The Large-scale Interstellar Medium of SS~433$/$W50 Revisited
}

\shorttitle{The Surrounding ISM of SS~433$/$W50}

\correspondingauthor{Yang Su}
\email{yangsu@pmo.ac.cn}

\author[0000-0002-0197-470X]{Yang Su}
\affil{Purple Mountain Observatory and Key Laboratory of Radio Astronomy,
Chinese Academy of Sciences, Nanjing 210034, China}

\author{Xin Zhou}
\affiliation{Purple Mountain Observatory and Key Laboratory of Radio Astronomy,
Chinese Academy of Sciences, Nanjing 210034, China}

\author{Ji Yang}
\affiliation{Purple Mountain Observatory and Key Laboratory of Radio Astronomy,
Chinese Academy of Sciences, Nanjing 210034, China}

\author{Yang Chen}
\affiliation{Department of Astronomy, Nanjing University,
Nanjing 210023, China}
\affiliation{Key Laboratory of Modern Astronomy and Astrophysics,
Nanjing University, Ministry of Education, Nanjing 210093, China}

\author{Xuepeng Chen}
\affiliation{Purple Mountain Observatory and Key Laboratory of Radio Astronomy,
Chinese Academy of Sciences, Nanjing 210034, China}

\author{Shaobo Zhang}
\affiliation{Purple Mountain Observatory and Key Laboratory of Radio Astronomy,
Chinese Academy of Sciences, Nanjing 210034, China}

\begin{abstract}
With new high-resolution CO and \mbox{H\,\textsc{i}} data, we revisited
the large-scale interstellar medium (ISM)
environment toward the SS~433$/$W50 system.
We find that two interesting molecular cloud (MC) concentrations, G39.315$-$1.155 and G40.331$-$4.302,
are well aligned along the precession cone of SS~433 within a smaller 
opening angle of $\sim \pm7^{\circ}$. The kinematic features of the two
MCs at $\sim$73--84~km~s$^{-1}$, as well as those of the corresponding atomic-gas counterparts, 
are consistent with the kinematic characteristics of SS~433.
That is, the receding gas from SS~433 jet is probably responsible for 
the redshifted feature of G39.315$-$1.155 near the Galactic plane
and the approaching one may power the blueshifted gas
of G40.331$-$4.302 toward the observer. 
Moreover, the \mbox{H\,\textsc{i}} emission at $V_{\rm LSR}\sim$70--90~km~s$^{-1}$
displays the morphological resemblance with the radio
nebula W50.
We suggest that the $V_{\rm LSR}$=77$\pm$5~km~s$^{-1}$ gas is physically
associated with SS~433$/$W50, leading to 
a near kinematic distance of 4.9$\pm$0.4~kpc for the system.
The observed gas features, which are located outside the current radio boundaries of W50, 
are probably the fossil record of jet--ISM interactions at $\sim 10^{5}$~years ago.
The energetic jets of the unique microquasar have profound effects on its
ISM environment, which may facilitate the formation of molecular gas
on the timescale of $\lsim$0.1~Myr for 
the ram pressure of $\sim 2\times10^6\ {\rm K\ cm^{-3}}$.
\end{abstract}


\section{INTRODUCTION}
The radio nebula W50, also known as supernova remnant 
(SNR) G39.7$-$2.0 \citep[e.g.,][]{2017yCat.7278....0G}, has 
a large angular extent of $\sim 120'\times60'$, which surrounds 
its central bright compact point source of SS~433. 
SS~433, a close massive binary system, is a Galactic microquasar 
consisting of a compact object and a massive donor star
\citep[e.g., see reviews of][]{1984ARA&A..22..507M,2004ASPRv..12....1F}.

The well-known and unique object SS~433 has been widely studied in 
mulit-wavelength observations, as well as theoretical analyses
and numerical simulations.
Most of these studies focused on the properties of the energetic
microquasar itself. On the other hand, many works were
also concentrated on the large-scale environment of the unusual system
(\citealp[e.g., radio continuum studies in][]{1980A&A....84..237G,1981A&A...103..277D,1986MNRAS.218..393D,1987AJ.....94.1633E,1998AJ....116.1842D,2011A&A...529A.159G,2018MNRAS.tmp...78B};
\citealp[IR studies in][]{1987PASP...99.1269B,1990A&A...240...98W,1996A&A...315L.113M};
\citealp[optical studies in][]{1980ApJ...236L..23V,1980MNRAS.192..731Z,1983ApJ...265..235M,2007MNRAS.381..308B,2010AN....331..412A};
\citealp[X-ray studies in][]{1983ApJ...273..688W,1994PASJ...46L.109Y,1996A&A...312..306B,1997ApJ...483..868S,1999ApJ...512..784S,2007A&A...463..611B};
\citealp[the multi-wavelength studies in][]{2005AdSpR..35.1062M};
\citealp[$\mbox{H\,\textsc{i}}$ and CO gas studies in][]{2007MNRAS.381..881L,2008PASJ...60..715Y};
\citealp[and the very recent magnetic fields and ionized gas studies in][]{2017MNRAS.467.4777F}).

Among the ample scope on the SS~433$/$W50 system, the interaction between
SS~433$/$W50 and its surrounding interstellar medium (ISM)
is a worthy topic for further studies.
The prominent phenomenon of the system is that
the elongation of W50 is exactly along the axis
of the precession cone of the SS~433 jets. The extension of
the relativistic jets of SS~433 (and/or jet counterparts in multiwavelength, e.g., X-ray, optical,
and radio emission) has an orientation coincident with that of W50
nebula, suggesting the physical connection between them.
But for all this, the origin of them is still uncertain and debated
\citep[e.g., more details in][]{2017MNRAS.467.4777F}.

In this paper, we use the Milky Way Imaging Scroll Painting
(MWISP\footnote{http://english.dlh.pmo.cas.cn/ic/}) CO data and
the complementary \mbox{H\,\textsc{i}} data from the Green Bank
Telescope (GBT) 100~m and Arecibo Observatory 305~m telescopes
to investigate the molecular and atomic gas properties of W50.
Thanks to the Galactic CO and \mbox{H\,\textsc{i}} surveys
with the high-dynamical range,
we can gain much insight into the ISM
environment of SS~433$/$W50 from the
combined high-quality molecular and atomic line observations.
Especially, the high-resolution (in spatial and velocity) CO and
\mbox{H\,\textsc{i}} data allow us to investigate the gas properties
and the kinematic features of the surrounding ISM of SS~433$/$W50 from the large-scale structures
of several degrees to the small-scale features of $\lsim 1$ arcmin.

\begin{figure}[ht]
\plotone{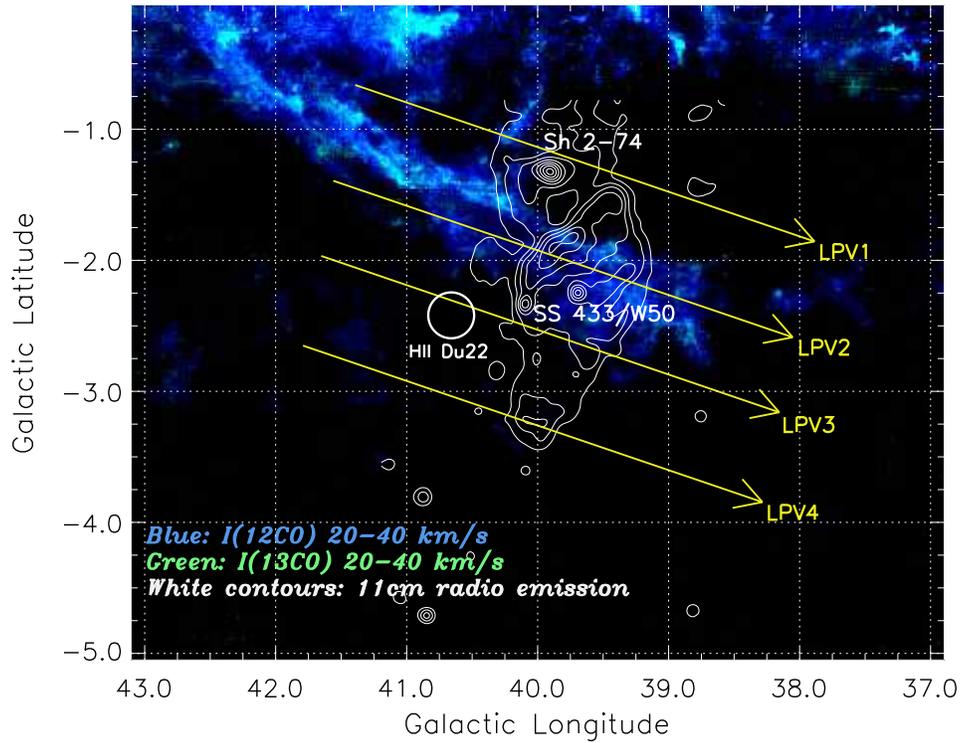}
\caption{The MWISP \twCO\ ($J$=1--0, blue) and \thCO\ ($J$=1--0, green)
intensity map toward SS~433$/$W50 in the 20--40~km~s$^{-1}$ interval,
overlaid with radio continuum contours
from the Effelsberg 11 cm survey \citep{1990A&AS...85..633R}.
The nearby sources of \HII\ regions Sh 2-74 and Du 22 are also labeled.
The four yellow arrows indicate the PV slices shown in Figure \ref{f2}.
\label{f1}}
\end{figure}

Throughout this paper, we use Galactic coordinates to identify directions on the sky,
which is convenient to analyze the structure of the gas on a large scale,
e.g., several degrees.
In particular, the well-known western and eastern jets$/$lobes of SS~433$/$W50
in Equatorial coordinates are described as the northwestern jet$/$lobe
(the receding one toward the Galactic plane)
and the southeastern jet$/$lobe (the approaching one away from the plane) 
in Galactic coordinates (see the radio morphology of
W50 in Figure \ref{f1} and the red and blue lines in Figure \ref{f4}), respectively.

\section{Observations and data}
\subsection{CO data}
The CO data used in this work are part of the MWISP project,
which is a large, unbiased, and high-sensitivity CO 
survey toward the Galactic plane for the region
of $l=[-10^{\circ}, 250^{\circ}]$ and $b=[-5^{\circ}, 5^{\circ}]$.
\twCO~($J$=1--0), \thCO~($J$=1--0), and C$^{18}$O~($J$=1--0) lines
were observed simultaneously using the 13.7~m 
millimeter-wavelength telescope located at Delingha in China. 
A $3\times3$ beam array \citep{Shan} was designed to monitor nine positions 
at once, increasing the mapping speed by roughly an order of magnitude. 
The total bandwidth is 1 GHz and the half-power beamwidth (HPBW)
of the telescope is about $50''$ for the three lines. 
The typical rms noise level is about 0.5 K for \twCO\ ($J$=1--0) at the
channel width of 0.16$\km\ps$ and 0.3 K for \thCO\ ($J$=1--0)
and C$^{18}$O ($J$=1--0) at 0.17$\km\ps$. 
The details of the 13.7 m telescope can be found from the status
report\footnote{http://www.radioast.nsdc.cn:81/mwisp.php}.

The observing strategy, the instrument, and the quality
of the CO observations were described in our recent 
paper \citep[see][]{2017ApJ...836..211S}. 
Briefly, Each $30'\times30'$ map was covered with position-switch On-The-Fly
mode at least
twice in scanning direction along Galactic longitude and 
latitude to minimize the fluctuation of noise.
The scan speed was $50''$s$^{-1}$ (or $75''$s$^{-1}$)
with a dump time of 0.3 s (or 0.2 s).
The sampling interval was $15''$ and the spacing between
scan rows was $10''$, fulfilling the requirements for oversampling
of the $50''$ beam of 13.7 m telescope.  
After the first order (linear) baseline fitting
and mosaicing the image, the final cube data
were constructed with a grid spacing of $30''$.
All data were reduced using the
GILDAS software\footnote{http://ascl.net/1305.010 or
http://www.iram.fr/IRAMFR/GILDAS}.
The SS~433$/$W50 region was completely mapped
during 2012 to 2015 \citep[see][]{2016ApJ...828...59S}.

\subsection{\mbox{H\,\textsc{i}} data and radio continuum emission}
To compare with the large-scale molecular gas, we used the 100~m
GBT 21~cm emission line of \mbox{H\,\textsc{i}}
\citep[see the details in][]{2007MNRAS.381..881L} as a tracer of the neutral atomic gas.
The final GBT \mbox{H\,\textsc{i}} data were gridded to 3\farcm5
with a velocity separation of 1.03$\km\ps$.
The further Galactic ALFA \mbox{H\,\textsc{i}} 
\citep[GALFA;][]{2011ApJS..194...20P} survey data, which were done 
with the Arecibo Observatory 305~m telescope, were also investigated.
The GALFA \mbox{H\,\textsc{i}} cube data
have a grid spacing of 1\farcm0 and a velocity channel separation 
of 0.184$\km\ps$. Typical noise levels are 0.1~K rms of brightness 
temperature in an integrated 1$\km\ps$ channel for both the GBT 
and GALFA \mbox{H\,\textsc{i}} data.
Finally, the radio continuum emission of SS~433$/$W50 was from
the Effelsberg 11 cm survey \citep{1990A&AS...85..633R}.

\section{The molecular and atomic gas toward SS~433$/$W50} 
Several works had been done to investigate the gas environment
toward the SS~433$/$W50 system
\citep[e.g.,][]{1983ApJ...272..609H,1998AJ....116.1842D,2000AdSpR..25..703D,2007MNRAS.381..881L,2008PASJ...60..715Y}.
However, conclusions from these studies, especially on the 
kinematic distance to SS~433$/$W50 (e.g., 2.2--5.5~kpc),
are not consistent with each other.
In these papers, the local standard of rest (LSR) velocity of the gas was
used to determine the
distance of the system based on \mbox{H\,\textsc{i}} absorption 
and/or CO$+$\mbox{H\,\textsc{i}} associations with W50 nebula.
As an association between SS~433$/$W50 and its surrounding gas
was established, the kinematic distance of them at a certain LSR
velocity can be determined from the Galactic rotation curve model.
Therefore, the key point to the argument is which of the gas components along
the line of sight (LOS) is actually associated with SS~433$/$W50.

A detailed analysis of the combined molecular and atomic gas
toward the direction could shed light on the connection between the 
SS~433$/$W50 system and its surrounding ISM. 
The new high-resolution and high-sensitivity CO observations, together with
the complementary \mbox{H\,\textsc{i}} data, may also provide us a good
opportunity to investigate interactions of the system 
with its ambient gas on a large scale.

Generally, a large amount of molecular gas traced by CO emission,
as well as the complicated atomic gas traced by \mbox{H\,\textsc{i}} data,
is seen in the region of SS~433$/$W50. 
For the convenience of discussion, the gas toward SS~433$/$W50 
was divided into four parts according to different 
velocities of $V_{\rm LSR}\lsim 0$~km~s$^{-1}$, $V_{\rm LSR}=$0--20~km~s$^{-1}$, 
$V_{\rm LSR}=$20--70~km~s$^{-1}$, and $V_{\rm LSR} \gsim$70~km~s$^{-1}$.
Position$-$velocity (PV) diagrams across the radio nebula W50
(e.g., see yellow arrows 
that are roughly perpendicular to the precession axis of SS~433 
in Figure \ref{f1}) were made in Figure \ref{f2} for subsequent analysis.

\subsection{Gas at $\lsim$0~km~s$^{-1}$}
The gas with a negative velocity is believed to lie beyond the solar circle
in the direction.
A little molecular gas with $V_{\rm LSR}\lsim 0$~km~s$^{-1}$ 
was detected near the radio boundary of SS~433$/$W50 
in the MWISP CO survey. 
These molecular clouds (MCs) are located within the distant Outer Arm 
\citep[e.g., MWISP G039.923$-$00.655 at $V_{\rm LSR}\sim -31$~km~s$^{-1}$,][]{2016ApJ...828...59S} 
or the Extreme Outer Galaxy region
\citep[e.g., MWISP G039.175$-$01.425 and MWISP G039.225$-$01.524 
at $V_{\rm LSR}\sim -55$~km~s$^{-1}$,][]{2017ApJS..230...17S}. 
We confirm that the MCs with negative velocities are not related
to SS~433$/$W50 due to their large kinematic distances,
e.g., $d>$13~kpc.

One may wonder whether these MCs were perturbed or
accelerated by the high-velocity gas from SS~433 jet. 
However, this possibility is excluded due to lacking 
any kinematic signatures (e.g., prominent line broadenings 
or asymmetric line profiles) and/or velocity gradients 
within the precession cone of SS~433 for the gas with 
negative velocities. 
After checking the channel maps of \mbox{H\,\textsc{i}} data,
we cannot find any large-scale morphological correspondence between 
the atomic gas and the extended radio emission of W50. 
There seems to be no significant accelerated gas 
toward us in the velocity range of $V_{\rm LSR}\lsim 0$~km~s$^{-1}$, 
which is in agreement with the analysis of \cite{2007MNRAS.381..881L}. 
We will mainly focus on the gas within the solar circle 
($V_{\rm LSR}\gsim 0$~km~s$^{-1}$) in the subsequent sections.

\subsection{Gas at 0--20~km~s$^{-1}$}
In the velocity range, most of the CO emission,
which is widely distributed in the field of view (FOV),
is obvious from the Local Arm because of their large-scale angular
sizes in space (e.g., several tens arcmins to several degrees) and
their broad distribution along the Galactic latitude 
(e.g., $b=-$5\fdg1 to $b=+$5\fdg1).
Briefly, the MCs, which are embedded in diffuse CO emission, 
display filamentary structures 
or irregular morphologies on a relatively large scale. 
Small and faint MCs with various structures
can be seen everywhere.
Some bright MCs with relatively small angular sizes (e.g., several arcmins
to $\lsim10$ arcmins), which are mainly concentrated in the Galactic plane
of $b \sim -$0\fdg4 -- +0\fdg6, probably belong to 
the distant Perseus Arm in the direction.
The details of the CO emission in the MWISP survey will be presented elsewhere.

According to the PV diagrams (Figure \ref{f2}) across the W50
nebula, we do not detect any protruding velocity structures that are often used to
trace shock--MC interactions
\citep[e.g.,][]{2014ApJ...788..122S,2014ApJ...796..122S,2017ApJ...845...48S}.
Further CO spectral analysis does not reveal any striking line 
diagnostics near and within the radio boundary of the nebula.
It indicates that the molecular gas in the interval of 0--20~km~s$^{-1}$
is relatively quiescent in the projected area of W50.
Wherever the molecular gas at such the velocity interval is from, 
there is little morphological correspondence 
between the molecular gas and the W50 nebula.
The \mbox{H\,\textsc{i}} emission, which is more diffuse 
and extended than CO emission, do not show any prominent 
connection with W50 either.
We conclude that the gas in the velocity range is not related to
SS~433$/$W50.

\subsection{Gas at 20--70~km~s$^{-1}$}
\subsubsection{The 170~pc long MC filament G40.82$-$1.41 at 29--34~$\km\ps$}
According to the CO intensity map in the interval of 20--40~km~s$^{-1}$
(Figure \ref{f1}), a giant molecular filament (GMF),
which is named as G40.82$-$1.41 from its geometric center
of the CO emission,
is revealed to extend about 5 degrees from ($l$=42\fdg9, $b=-$0\fdg1) to
($l$=38\fdg7, $b=-$2\fdg7).
The distance to the GMF is estimated to be 1.6--2.0 kpc
\citep[e.g., the 3D extinction map from][]{2015ApJ...810...25G}
or $\sim 1.8$--2.1 kpc from the CO peak velocities of $V_{\rm LSR}\sim 29$--$34\km\ps$ 
\citep[the near kinematic distance from the A5 model of][]{2014ApJ...783..130R}.
The length of the GMF is thus about 170~pc at a distance of $\sim$2.0~kpc.

The GMF G40.82$-$1.41 is a clustering of tens of small filaments with
somewhat different LSR velocities. These small filaments 
consisting of multiple components exhibit complicated structures. 
Many of them are oriented parallel to the long axis of the GMF
and appear to be twisted between each other. 
The southwestern part of the GMF across the radio nebula W50
was suggested to have a connection with the SNR W50
\cite[see Figure 3 and discussions in][]{1983ApJ...272..609H}.
Based on PV diagrams (e.g., LPV2 and LPV3 in Figure \ref{f2}), however,
we cannot find significant kinematic features of shock--MCs interactions
near and within the boundary of W50 nebula.
Any asymmetric CO line profiles, which can be seen
everywhere both within and outside the W50 nebula, are 
obviously attributed to overlapping components at somewhat different velocities.

\subsubsection{The MCs along the precession cone of SS~433 
and the corresponding \mbox{H\,\textsc{i}} gas at $\sim$40--60~km~s$^{-1}$}
Several interesting MCs seem roughly aligned along the precession cone of SS~433, 
which was suggested to be the evidence of SS~433$/$W50--ISM interactions
\citep[see Figures 1--4 and discussions in][]{2008PASJ...60..715Y}.
Combined with the radio continuum and \mbox{H\,\textsc{i}} studies 
by \cite{1998AJ....116.1842D},  \cite{2008PASJ...60..715Y} then placed 
these MCs (the northern MCs of N1--N4 at $V_{\rm LSR}\sim 53\km\ps$ 
and the southern MCs of S1--S6 at $V_{\rm LSR}\sim 43\km\ps$, see Table 1 in their paper) 
and the associated SS~433 at a near kinematic distance of $\sim$3~kpc.

Figure \ref{f3} shows the $WISE$ 22~$\mu$m emission overlaid with MWISP
CO contours \citep[blue, corresponding to MCs S1--S6 in][]{2008PASJ...60..715Y}
in the interval of 39--51$\km\ps$ and GALFA \mbox{H\,\textsc{i}}
contours (red) in the interval of 35--48$\km\ps$.
Obviously, the bright diffuse IR emission, which displays multiple
shell-like or bubble-like features, is coincident well with the gas emission at
$V_{\rm LSR}\sim 43\km\ps$, suggestive of an association between the
IR features and the gas. 
The near kinematic distance of the gas is about 2.7 kpc
\citep{2014ApJ...783..130R}, which is in good agreement with the
2.0---2.5~kpc from the 3D extinction map of \cite{2015ApJ...810...25G}.
The atomic and molecular gas, 
together with the bright thermal dust association
centered at ($l\sim$40\fdg4, $b\sim-$4\fdg3) with a radius of $\sim 1^{\circ}$,
is probably related to the star-forming regions at a distance of $\sim$2.5~kpc 
(or 190~pc below the Galactic plane).
The total mass of the atomic$+$molecular gas within the region
of $\sim 90 \times 90$~pc$^{2}$ is $\gsim 1 \times 10^{5} \Msun$.
It is hard to believe that such amounts of gas can originate from 
the jet--ISM process of SS~433.

On the other hand, MCs at 50--55$\km\ps$ were suggested to be
related to the W50 lobe toward the Galactic plane due to
the spatial coincidence between them \citep{2000AdSpR..25..703D,
2001A&A...366.1035C,2008PASJ...60..715Y}.
Most of the MCs, which are near the \mbox{H\,\textsc{ii}} region Sh 2-74
and close to the Galactic plane,
also have corresponding bright IR emission. 
We argue that these MCs with prominent \thCO\ emission are probably associated with 
\mbox{H\,\textsc{ii}} region Sh 2-74 and/or belong to the nearby 
MC complex with active star formation (see Section 3.3.3),
which agrees with suggestions from previous studies 
\citep[e.g.,][]{1989ApJ...338..945B,2005AdSpR..35.1062M}.

At 38--44$\km\ps$, a large \mbox{H\,\textsc{i}} void-like feature 
roughly shows the morphological correspondence with the radio 
emission of W50 \citep[e.g., see Figure~14 in][]{2007MNRAS.381..881L}.
However, the large-scale atomic cavity seen in the integrated map
seems to be comprised of several--separated substructure of 
\mbox{H\,\textsc{i}} emission. We cannot find 
fine morphological agreement between the radio continuum emission of W50 
and its nearby atomic gas 
after analyzing the Arecibo \mbox{H\,\textsc{i}} data channel-by-channel.
We agree with the conclusion of \cite{2007MNRAS.381..881L} that
the gas in the velocity range is not physically associated with SS~433$/$W50.

\subsubsection{Other interesting MCs}
\begin{figure}
\plotone{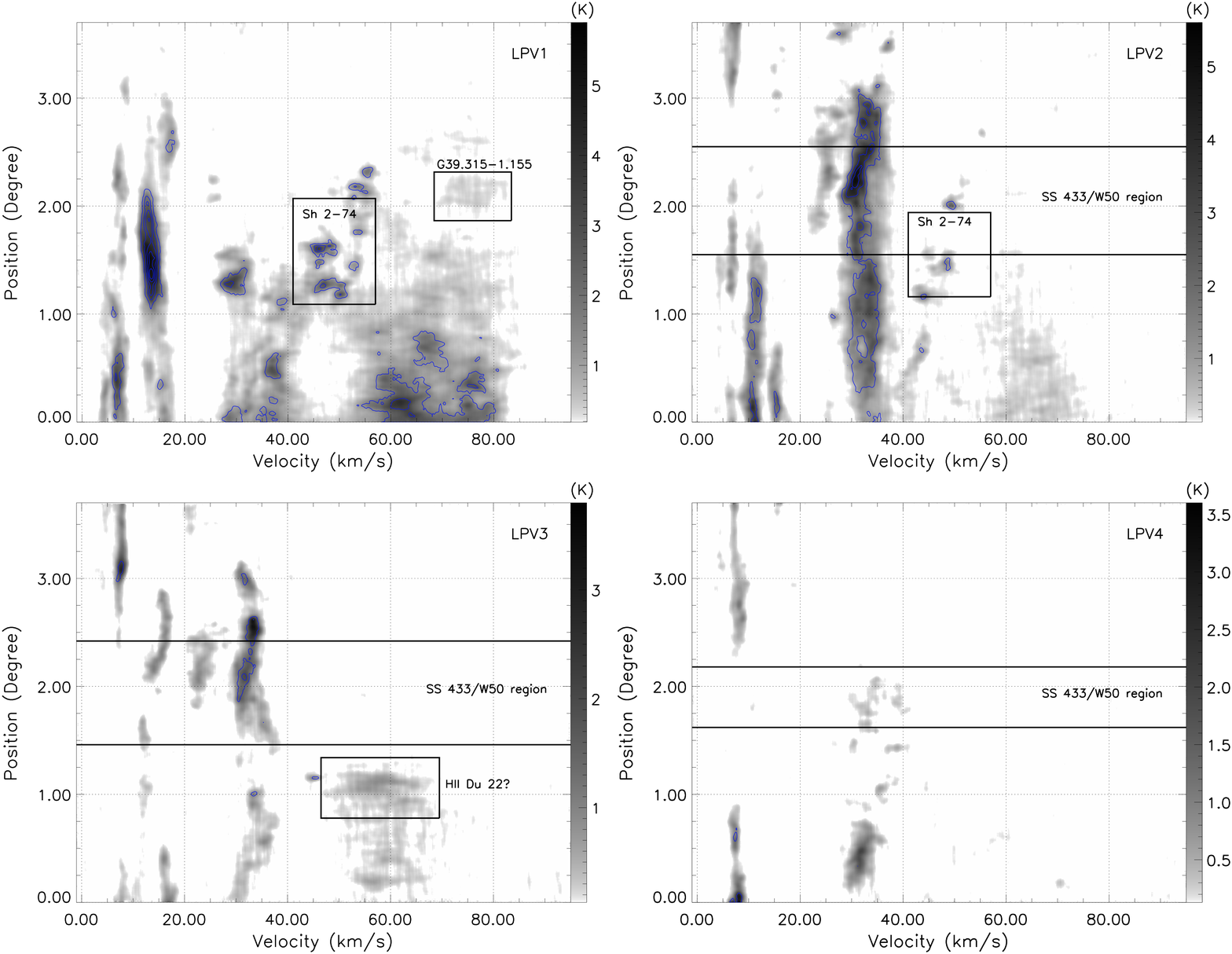}
\caption{ 
PV diagrams of \twCO\ ($J$=1--0) emission along LPV1, LPV2, LPV3, and LPV4,
overlaid with the blue contours of \thCO\ emission. The overlaid contour
levels start from 0.2 K and increase by a step of 0.4 K.
The PV slices have a length of 3\fdg7
(($l=$41\fdg393, $b=-$0\fdg662) to ($l=$37\fdg886, $b=-$1\fdg856) for LPV1,
 ($l=$41\fdg560, $b=-$1\fdg393) to ($l=$38\fdg053, $b=-$2\fdg587) for LPV2,
 ($l=$41\fdg652, $b=-$1\fdg967) to ($l=$38\fdg155, $b=-$3\fdg160) for LPV3, and
 ($l=$41\fdg791, $b=-$2\fdg651) to ($l=$38\fdg284, $b=-$3\fdg845) for LPV4)
and a width of 0\fdg508.
The horizontal lines show the radio regions of W50 nebula.
\label{f2}}
\end{figure}

\begin{figure}
\plotone{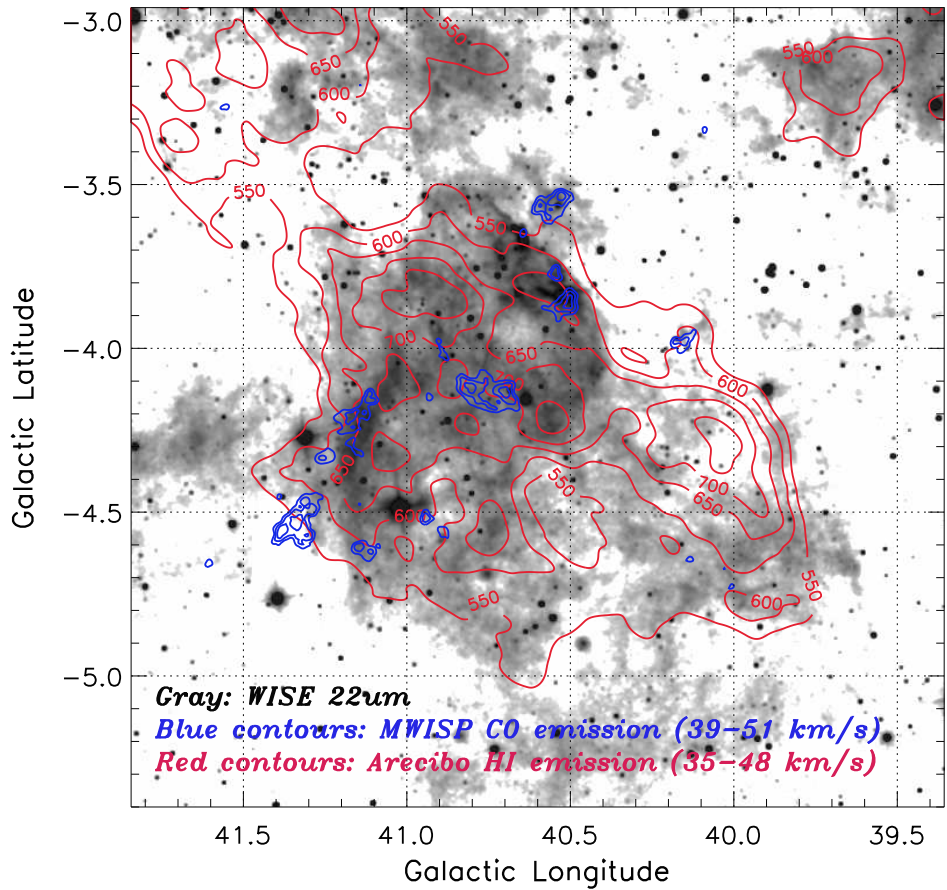}
\caption{
$WISE$ 22~$\mu$m emission toward ($l$=40\fdg4, $b=-$4\fdg3),
overlaid with molecular gas emission (blue contours with 3, 6, and 9~K~km~s$^{-1}$ 
for the MWISP \twCO\ in the interval of 39--51~km~s$^{-1}$) 
and atomic gas emission (red contours with 550, 600, 650, 700, and 750~K~km~s$^{-1}$
for the GALFA \mbox{H\,\textsc{i}} in the interval of 35--48~km~s$^{-1}$).
\label{f3}}
\end{figure}

The \mbox{H\,\textsc{ii}} region Sh 2-74 ($l\sim$39\fdg9, $b\sim-$1\fdg3), 
which has a large radio size of $\sim 1^{\circ} \times 1^{\circ}$ at a distance 
of about 3~kpc \citep[e.g.,][]{1987AJ.....94.1633E,2003A&A...397..213P},
lies near the northeastern edge of the W50 nebula (Figure \ref{f1}).
Using MWISP CO data,
we find that the molecular gas at 40--55$\km\ps$
\citep[$V_{\rm peak}\sim 48\km\ps$, which agrees well with the RRL central velocity 
of $\sim 47.7\km\ps$ for Sh 2-74,][]{2012MNRAS.422.2429A},
is physically associated with the 
\mbox{H\,\textsc{ii}} region according to the morphological 
agreement and the corresponding kinematic features between 
the ionized gas of the \mbox{H\,\textsc{ii}} region and 
the surrounding MCs (see LPV1 and LPV2 in Figure \ref{f2}).

An interesting MC at ($l\sim$40\fdg66, $b\sim-$2\fdg42, $V_{\rm peak}\sim 57\km\ps$) 
has broad CO wings in the velocity range of 45--72$\km\ps$ 
(see LPV3 in Figure \ref{f2}), which is very likely associated with
the \mbox{H\,\textsc{ii}} region Du 22 centered at ($l=$40\fdg6567, $b=-$2\fdg4658)
\citep{1976A&AS...25...25D}. 
The near kinematic distance of the object and
the associated MCs is estimated to be $\sim$3.5~kpc. The far one of $\sim$9.0~kpc
may be excluded due to the unreasonable scale-height of $\sim$400~pc.

Based on the above analysis,
we argue that the MCs in the velocity range 
of 20--70$\km\ps$ are accidental superpositions along
the LOS and are not associated with the  SS~433$/$W50 system.
Both of the very long GMF
\citep[see discussions in][]{1983ApJ...272..609H} at $V_{\rm LSR}\sim$~29--34$\km\ps$
and the gas features at $V_{\rm LSR}\sim$~40--60$\km\ps$
\citep{1998AJ....116.1842D,2000AdSpR..25..703D,2008PASJ...60..715Y} 
appear to be just the foreground gas of SS~433$/$W50 in the FOV.
We will not discuss these MCs
any further because the above gas is not related to SS~433$/$W50 at a
distance of 4.9~kpc (see Section 4).

\subsection{Gas at $\gsim$ 70~km~s$^{-1}$}
\begin{figure}
\plotone{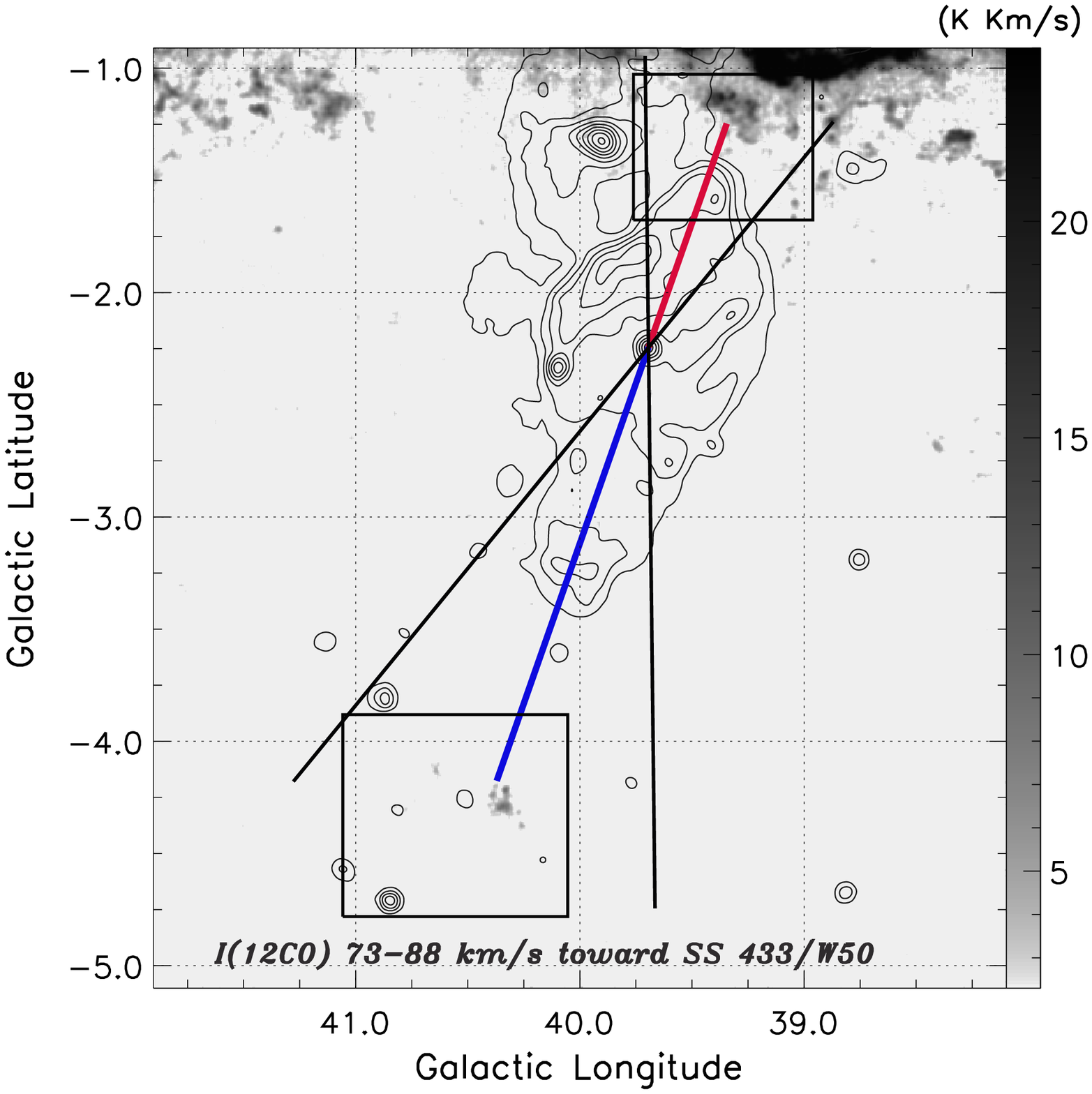}
\caption{Integrated \twCO\ ($J$=1--0) emission toward SS~433$/$W50
in the interval of 73--88~km~s$^{-1}$, overlaid with the same radio contours as in
Figure \ref{f1}.
The thick red and blue line indicates the precession axis 
of the SS~433 jets.
\citep[see e.g.,][]{1981ApJ...246L.141H,2002MNRAS.337..657S}.
The black lines indicate the cone-opening angle of $\pm 20^{\circ}$ 
around the precession axis.
The black boxes indicate the two regions shown in Figure \ref{f5}.
\label{f4}}
\end{figure}

There are only a few CO clouds ($V_{\rm LSR} \gsim$ 70~km~s$^{-1}$, 
$T_{\rm peak}\sim$ 1--2~K, and size$~\sim$1--2 arcmin$^{2}$)
within the radio boundary of the W50 nebula, 
e.g., ($l=$39\fdg225, $b=-$1\fdg942, $V_{\rm peak}\sim 73\km\ps$)
and ($l=$39\fdg692, $b=-$2\fdg450, $V_{\rm peak}\sim 74\km\ps$),
indicating that the CO emission at such velocities 
is very weak. Interestingly, two intriguing MCs,
which are named as G39.315$-$1.155 and G40.331$-$4.302,
are exactly aligned along the precession axis of
SS~433 (see two small boxes in Figure \ref{f4}). 

\begin{figure}
\gridline{\fig{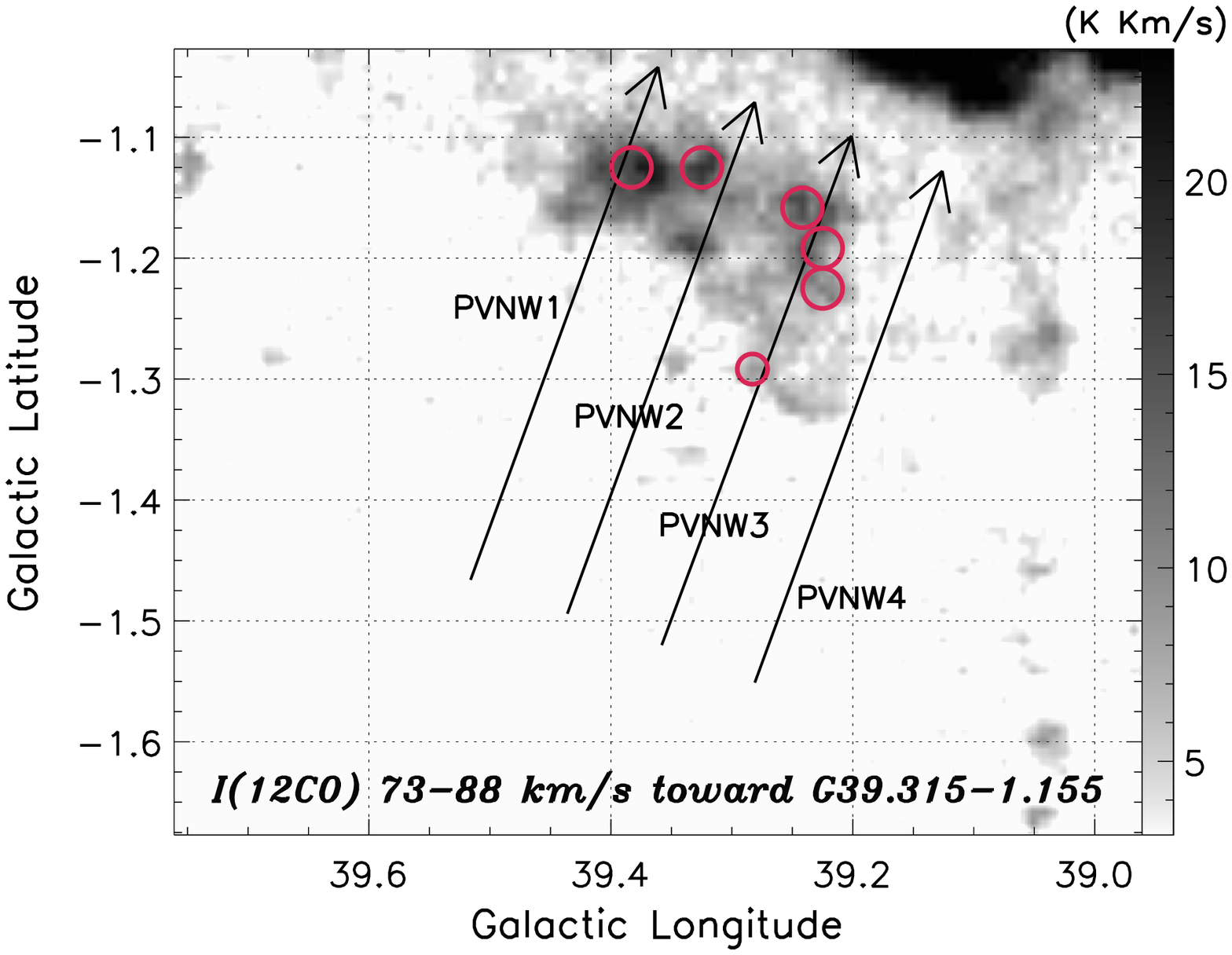}{0.55\textwidth}{}
          \fig{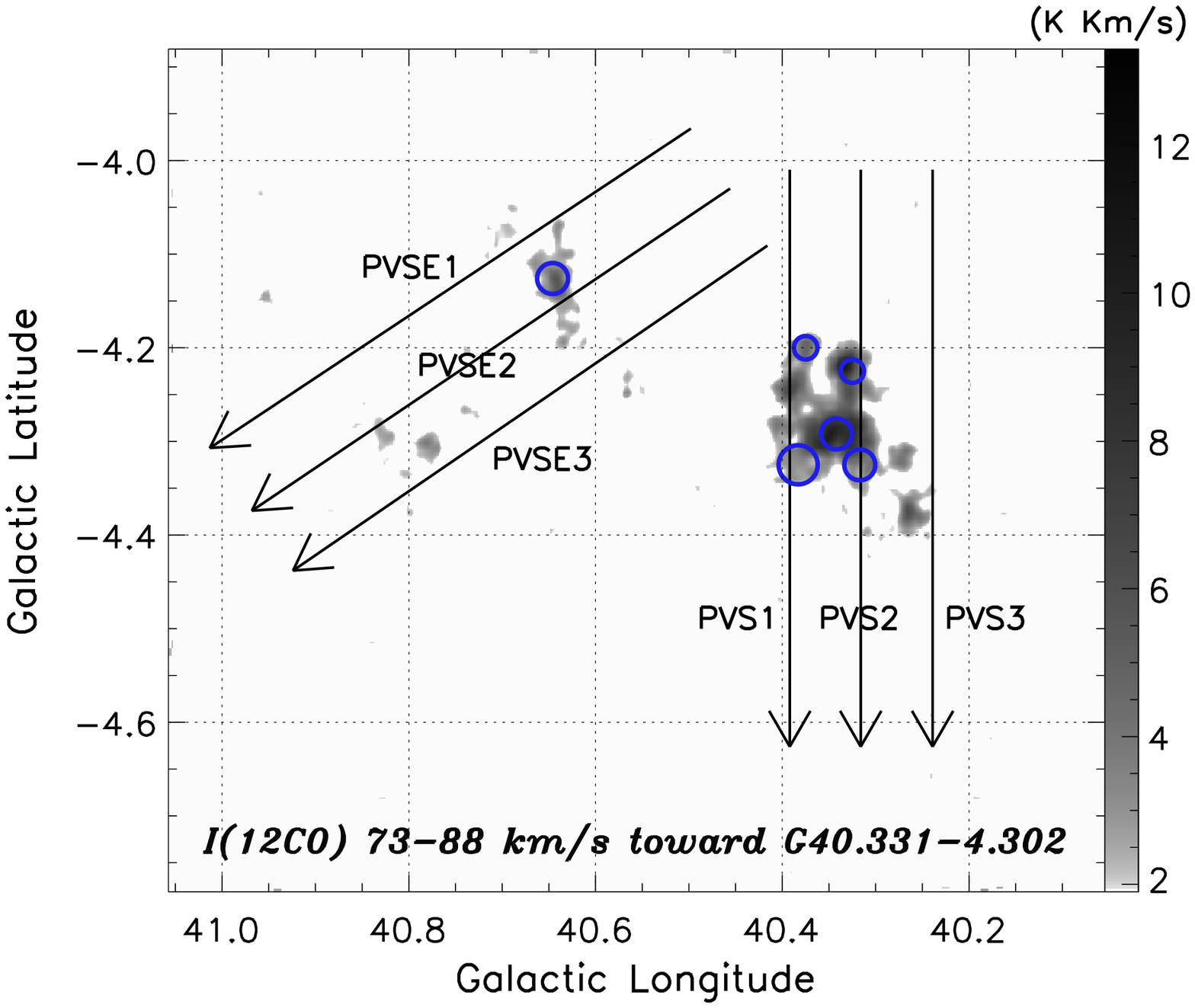}{0.55\textwidth}{}
          }
\caption{
Left panel: \twCO\ ($J$=1--0) emission toward the northwestern region
(MC G39.315$-$1.155) of SS~433$/$W50.
The four arrows indicate the PV slices shown in Figure \ref{f6}.
The six red circles indicate positions of the typical spectra shown
in Figure \ref{f8}.
Right panel: \twCO\ ($J$=1--0) emission toward the southeastern region
(MC G40.331$-$4.302) of SS~433$/$W50.
The six arrows indicate the PV slices shown in Figure \ref{f7}.
The six blue circles indicate positions of the typical spectra shown
in Figure \ref{f9}. 
\label{f5}}
\end{figure}

Figure \ref{f5} shows a close-up view of the MCs
G39.315$-$1.155 and G40.331$-$4.302.
The northwestern MC G39.315$-$1.155 appears to display a coherent structure
in the CO intensity map of 73--88~km~s$^{-1}$, while the 
southeastern MC G40.331$-$4.302 consists of two parts, 
the faint CO emission in the eastern region (near PVSE lines in the right panel
of Figure \ref{f5}) 
and the relatively strong emission in the western region
(near PVS lines in the figure).
Both of the two MC concentrations are along the direction from the northeast to 
the southwest, which are nearly perpendicular to 
the precession axis of SS~433 (see red and blue lines in Figure \ref{f4}).

PV diagrams along selected lines (see black arrows in Figure \ref{f5})
are shown in Figures \ref{f6} and \ref{f7}, for MCs G39.315$-$1.155 and 
G40.331$-$4.302, respectively. 
Typical spectra of some regions (see circles in Figure \ref{f5}) 
are shown in Figures \ref{f8} and \ref{f9}, respectively.
It is interesting to note that MC G39.315$-$1.155 displays
redshifted features while MC G40.331$-$4.302 displays
somewhat blueshifted features in both of the PV diagrams
and the typical spectra.

Table~1 lists properties of the atomic and molecular gas toward the
main part of G39.315$-$1.155 and G40.331$-$4.302.
The column density of atomic gas of G40.331$-$4.302 is 
calculated via the conversion factor of $1.823\E{18}$~cm$^{-2}$(K~km~s$^{-1})^{-1}$
\citep{1990ARA&A..28..215D}.
We cannot calculate the atomic-gas properties of G39.315$-$1.155
due to the strong background \mbox{H\,\textsc{i}} emission
near the Galactic gas plane.
The peak temperature of MC G40.331$-$4.302 is $\sim$3.4~K and $\sim$0.6~K
for \twCO\ and \thCO, respectively.
Note that if we use LTE assumption, an excitation temperature of 10 K,
the beam filling factor of $\sim$0.5,
and $N$(H$_2$)/$N(^{13}$CO) $\approx7\E{5}$, the H$_2$ column density of
G40.331$-$4.302 is about 1.1$\times10^{21}$~cm$^{-2}$, which is somewhat larger than
the estimated value from the X-factor method
\citep[e.g., the mean CO-to-H$_2$ mass
conversion factor of $2\E{20}$~cm$^{-2}$(K~km~s$^{-1})^{-1}$,]
[]{2001ApJ...547..792D,2013ARA&A..51..207B}. It can be naturally explained
because we adopt the peak emission of \twCO\ and \thCO\ to represent
the property of the whole MC.

For the gas in G40.331$-$4.302,
the velocity dispersion ($\sigma_{v}=\frac{\Delta V_{\rm FWHM}}{2.355}$) 
is estimated to be $\sim$6~km~s$^{-1}$ for the atomic gas
and $\sim$1.7~km~s$^{-1}$ for the molecular gas, respectively.
The mean density of the initial gas environment 
is on the order of $\sim 1$~cm$^{-3}$, which is a reasonable value 
estimated from the $V_{\rm LSR}\sim$70--90~km~s$^{-1}$
\mbox{H\,\textsc{i}} emission along the precession axis of SS~433.
For MC G39.315$-$1.155, the velocity dispersion of the
molecular gas is roughly 3--4~km~s$^{-1}$, which is approximately two times
larger than that of MC G40.331$-$4.302.
The property indicates that MC G39.315$-$1.155 
is much more turbulent than MC G40.331$-$4.302.

For G40.331$-$4.302, the molecular gas is mainly concentrated
in a slab with a size of $\sim 32' \times 8'$ (e.g., Figures \ref{f10} and \ref{f12}).
The total mass of G40.331$-$4.302 should be somewhat larger
than the estimated value of $\sim3\times 10^{3} \Msun$ in Table~1 because of the unaccounted
gas in the northeastern region of the cloud (near PVSE in Figure 
\ref{f5}) and the potential H$_2$ gas in CO-dark regions (see Section 6).
For G39.315$-$1.155, the H$_2$ mass within a size of $\sim 17' \times 8'$
is $\sim1.1\times 10^{4} \Msun$.

Obviously, the total mass of G39.315$-$1.155 is larger than that of
G40.331$-$4.302.
The two MCs also have different angular distances 
from SS~433 (e.g., $\sim 69'$ for G39.315$-$1.155 vs.
$\sim 129'$ for G40.331$-$4.302) and different extension
along the northeast--southwest direction
(e.g., $\sim 17'$ for G39.315$-$1.155 vs. $\sim 32'$ for G40.331$-$4.302).
Actually, the northwestern radio lobe of W50 is famously shorter and brighter
than the southeastern one, which is widely attributed to the
denser ambient medium close to the Galactic plane 
\citep[e.g.,][]{1998AJ....116.1842D,2007MNRAS.381..881L,2011MNRAS.414.2838G}.
The two MCs features presented here can also be explained by the 
same reason of the denser environment near the Galactic plane,
if these MCs are governed by the SS~433 jets/outflows (see Sections 4--6).

The extended and fragmented CO gas
of the two MC concentrations is well confined to a cone
with a small angular extent of $(\frac{17}{69} \times \frac{180}{\pi})_{\rm G39.315} 
\sim (\frac{32}{129} \times \frac{180}{\pi})_{\rm G40.331} \approx 14^{\circ}$ 
with respect to SS~433.
We note that such the angular extent of the molecular gas
is comparable to the result
from the current X-ray jet of SS~433
\citep[e.g., $\sim 18^{\circ}$, also seen from SS~433,][]{2007A&A...463..611B}.

To search for possible evidence of the W50--ISM association, 
we constructed four integrated \mbox{H\,\textsc{i}} maps in the intervals
of 85--88, 82--85, 79--82, and 76--79~km~s$^{-1}$ (Figure \ref{f11}).
Several \mbox{H\,\textsc{i}} features
at the velocity range are indeed found to be positionally coincident with 
the bright radio shell of W50 in the FOV. 
Firstly, the \mbox{H\,\textsc{i}} maps exhibit a cavity-like structure
near the Galactic plane, 
which coincides with the radio morphology of the W50 nebula
\citep[also see the static ring in][]{2007MNRAS.381..881L}.
Secondly, several features with enhanced \mbox{H\,\textsc{i}}
emission match the bright radio shells of W50 very well
(e.g., see panels c and d in Figure \ref{f11}),
Thirdly, these \mbox{H\,\textsc{i}} enhancements, as well as the bright
radio shells, roughly extend along the northwest--southeast direction,
which is consistent with the trend of the precession axis of SS~433
(note that the axis is roughly perpendicular to the LOS 
with an inclination angle of $\approx 80^{\circ}$, \citealp{2001ApJ...561.1027E}).

Finally, the \mbox{H\,\textsc{i}} gas at
$V_{\rm LSR}\sim$70--90~km~s$^{-1}$ is associated with MC G40.331$-$4.302
(Figures \ref{f10}--\ref{f12}).
The \mbox{H\,\textsc{i}} concentration near MC G40.331$-$4.302, 
which seems to be a part of an expanding \mbox{H\,\textsc{i}} shell
near the southeastern radio lobe of W50
(or the outermost parts of the approaching gas from the SS~433 jet), 
was suggested to be related to
SS~433$/$W50 \citep{2007MNRAS.381..881L}.
Using the Arecibo \mbox{H\,\textsc{i}} data, we find that 
the atomic gas, as well as the corresponding
molecular gas at the velocity (e.g., the main part of MC G40.331$-$4.302),
is indeed approaching the observer (Figures \ref{f11} and \ref{f12}), 
indicative of the association between the gas and the SS~433 jet.

\section{Association of SS~433$/$W50 with the $\sim$77~km~s$^{-1}$ clouds and the distance}
Figure \ref{f4} displays that two MC concentrations exactly lie
projected on the precession axis of the SS~433 jets. Further analysis shows
that the kinematic features of the two MC concentrations are
consistent well with the behavior of the SS~433 jets.
That is, MC G39.315$-$1.155 displays the redshifted feature 
(e.g., especially, PVNW3 and PVNW4 in Figure \ref{f6})
corresponding to the receding jet of SS~433
and MC G40.331$-$4.302 displays the blueshifted feature
(e.g., especially, PVS1 and PVS2 in Figure \ref{f7})
corresponding to the approaching one.

The blueshifted feature of MC G40.331$-$4.302
is clearly confirmed from the accompanying atomic gas
using the Arecibo \mbox{H\,\textsc{i}} data (see Figure \ref{f12}). 
We speculate that MC G39.315$-$1.155 also has its
accompanying receding atomic gas. 
However, the very strong background \mbox{H\,\textsc{i}} emission
near the Galactic plane prevents us from obtaining 
its detailed kinematic information.
It is very difficult
to discern the possible kinematic feature at levels of
several K from the strong background emission at several tens K.
We also emphasize that we cannot identify the disturbed molecular gas
in the region closer to the Galactic plane (e.g., $b \gsim -$1\fdg0)
due to the complicated CO emission (e.g., strong \thCO\ emission and multiple
CO peaks in the velocity interval of 70--85~km~s$^{-1}$).

But despite all this, a \mbox{H\,\textsc{i}} cavity, which was
identified by \cite{2007MNRAS.381..881L} using the GBT observations,
does appear toward the W50 radio lobe near the Galactic plane 
using the new Arecibo data (Figure \ref{f11}).
MC G39.315$-$1.155 is roughly touching the top wall of the \mbox{H\,\textsc{i}}
cavity (e.g., red contours in panel d of Figure \ref{f11}).
Actually, a patch of atomic gas with enhanced \mbox{H\,\textsc{i}} emission 
at $\sim$80--90~km~s$^{-1}$
(see the \mbox{H\,\textsc{i}} gas near red contours in Figure \ref{f11})
seems to be coincident with the MC, suggesting the possible atomic gas
away from the observer.

Additionally, the \mbox{H\,\textsc{i}} emission at $\sim$70-90~km~s$^{-1}$
displays compelling morphological evidence for an association between W50 and the
atomic gas (e.g., Figure \ref{f11}).
Combined the above analysis and results from 
Section 3.4, all the evidence points to the gas at $\sim$70--90~km~s$^{-1}$.
We thus argue that the $\sim$77~km~s$^{-1}$ gas is physically
associated with SS~433$/$W50, which agrees well with the 
previous study by \cite{2007MNRAS.381..881L}.

We note that the heliocentric systemic radial velocity of 56$\pm$2~km~s$^{-1}$
was suggested to be related to SS~433$/$W50 from deep
optical observations toward the whole system \citep{2007MNRAS.381..308B}.
The heliocentric radial velocity of $\sim$56~km~s$^{-1}$ can be transformed
to a LSR velocity of $\sim$75~km~s$^{-1}$ \citep[e.g., see][]{2009ApJ...700..137R,
2014ApJ...783..130R}, which is in excellent agreement with our
finding of $V_{\rm LSR}=$77$\pm5$~km~s$^{-1}$ for the whole system
on a large scale.

Accordingly, the LSR velocity of 77$\pm$5~km~s$^{-1}$,
where the $\pm5$~km~s$^{-1}$ is the velocity error for possible
peculiar motions, corresponds to a near kinematic distance of 4.9$\pm$0.4~kpc
\citep[e.g., the A5 model in][]{2014ApJ...783..130R}.
The far kinematic distance is excluded due to lacking \mbox{H\,\textsc{i}}
absorption at the velocity near the tangent point 
\citep[e.g., $V_{\rm LSR}\sim$85~km~s$^{-1}$ or 
$d_{\rm tangent}\sim$6.4 kpc, see Figures 3 and 4 in][]{2007MNRAS.381..881L}.

We find that our new kinematic distance of 4.9$\pm$0.4~kpc is somewhat smaller 
than the value of 5.5$\pm$0.2~kpc from \cite{2007MNRAS.381..881L}. The 
discrepancy of the two estimates comes from the different parameters of the 
Galactic rotation curve model. In the work of \cite{2007MNRAS.381..881L},
they used a flat rotaion curve with $R_0$=8.5~kpc (distance of Sun from
Galactic Center) and $V_0$=220~km~s$^{-1}$ (rotation speed of Galaxy at $R_0$).
Accordingly, the LSR velocity of 75$\pm$6~km~s$^{-1}$ from the atomic
gas yielded the kinematic distance of 5.5$\pm$0.2~kpc. The velocity uncertainty
of $\pm$6~km~s$^{-1}$ 
in their work is the typical random motion of cool \mbox{H\,\textsc{i}} clouds.
In our case, we used the new Galactic rotation curve model,
in which the values of $R_0$=8.34~kpc and $V_0$=240~km~s$^{-1}$ are adopted
\citep[see Table 4 in][]{2014ApJ...783..130R}.
Thus, the 77$\pm$5~km~s$^{-1}$ from the CO gas leads to a distance of 4.9$\pm$0.4~kpc.
The velocity uncertainty of $\pm$5~km~s$^{-1}$,
which is roughly comparable to the value of $\pm$10~km~s$^{-1}$ for the 
typical peculiar motions of the high mass star forming regions
\citep{2014ApJ...783..130R}, is from the LSR 
velocity difference of the two MCs G39.315$-$1.155 and G40.331$-$4.302
(see Figures 6--9).

Finally, the distance to SS 433 is estimated
to be about 4.5--5.5~kpc \citep[depending on the different authors 
with somewhat different considerations, e.g.,][]{1981ApJ...246L.141H,
1993A&A...270..177V,2002MNRAS.337..657S,2004ApJ...616L.159B,2013ApJ...775...75M,2014A&A...562A.130P}
based on the kinematic model of the proper motions of the SS 433 jets.
Our near kinematic distance of 4.9$\pm$0.4~kpc is 
consistent well with the above estimates from
the traditional kinematic model originally pioneered by \cite{1981ApJ...246L.141H}.

We construct a cartoon to elucidate the association between
the observed CO$+$\mbox{H\,\textsc{i}} features and the SS~433$/$W50 system
(Figure \ref{f13}).
The main features of the ISM surrounding SS~433$/$W50, such as
the \mbox{H\,\textsc{i}} cavity near the Galactic plane,
the \mbox{H\,\textsc{i}} wall toward the approaching gas from the SS~433 jet,
and the corresponding kinematic characteristics of the gas and the jets,
are all included in the map.

In Figure 13, the opening angle from CO data is about $\pm10^{\circ}$,
which is comparable to the results from the radio and X-ray studies of W50 at large distances 
\citep[e.g., jet--ISM encounters at $\sim$35-80 arcmin from SS~433,][]
{1996A&A...312..306B,1998AJ....116.1842D,2007A&A...463..611B},
but much smaller than the $\pm20^{\circ}$ of the current precession cone of the SS~433 jets 
at the inner region \citep[e.g., $\lsim$6 arcsec from the 
compact source,][]{1981ApJ...246L.141H,2002MNRAS.337..657S}.
The discrepancy of the two opening angles with respect to SS~433 
is possibly due to the changing state of the jet precession with time
\citep[e.g.,][]{1990ApJ...350..561K,2008MNRAS.387..839Z,2011MNRAS.414.2838G}
and/or some hydrodynamical recollimated mechanisms for a precessing jet
\citep[e.g.,][]{1983ApJ...272...48E,1993ApJ...417..170P,2015A&A...574A.143M}.

Whatever the exact scenarios, the violent jet--ISM interactions seem to be
taking place around the precession axis of SS 433 with a small opening angle.
The bulk kinetic energy and momentum of the high-velocity gas will 
produce the high-pressure environment, leading to the rapid H$_2$ formation 
in such interaction regions (see Section 6). As a result, the new-formed 
molecular gas is likely distributed around an area near the outermost 
parts of the jet cones with a limited opening angle.
On the other hand, the SS 433's jets can shock the ISM and produce the
over-pressured environment with respect to the
surrounding gas, which may inflate a bubble propagating away
from the central source. 
The cavity-like structures of the atomic gas 
are probably such a case (e.g., the roughly ``eight-shaped" morphology of the atomic gas in
panel d of Figure \ref{f11}).
The size of the upper cavity-like feature of the atomic gas is much smaller than
that of the lower one, which is also probably due to the 
relatively higher density environment close to the Galactic plane.

\section{The fossil record of jet--ISM interactions}
SS~433 shows prominent jet activity, which deposits amounts of kinetic
energy into its surrounding ISM from several tens astronomical units to
dozens parsecs \citep[e.g., see Figure 8 in][]{2004ASPRv..12....1F}.
Now that the energetic jets of SS~433 may leave some mark on its ambient ISM,
it is interesting to search for some corresponding counterparts of
jets--ISM interactions. Several works has been done in earlier studies
using radio \citep[e.g.,][]{1981ApJ...246L.141H,1986MNRAS.218..393D,1998AJ....116.1842D,
2004ApJ...616L.159B,2007MNRAS.381..881L},
X-ray \citep[e.g.,][]{1983ApJ...273..688W,1996A&A...312..306B,
1997ApJ...483..868S,2007A&A...463..611B},
and optical \citep[e.g.,][]{1980MNRAS.192..731Z,2007MNRAS.381..308B} observations.
In this Section, we propose that the energetic microquasar SS~433 probably has
a more profound effect on its ambient ISM. 

\begin{figure}
\plotone{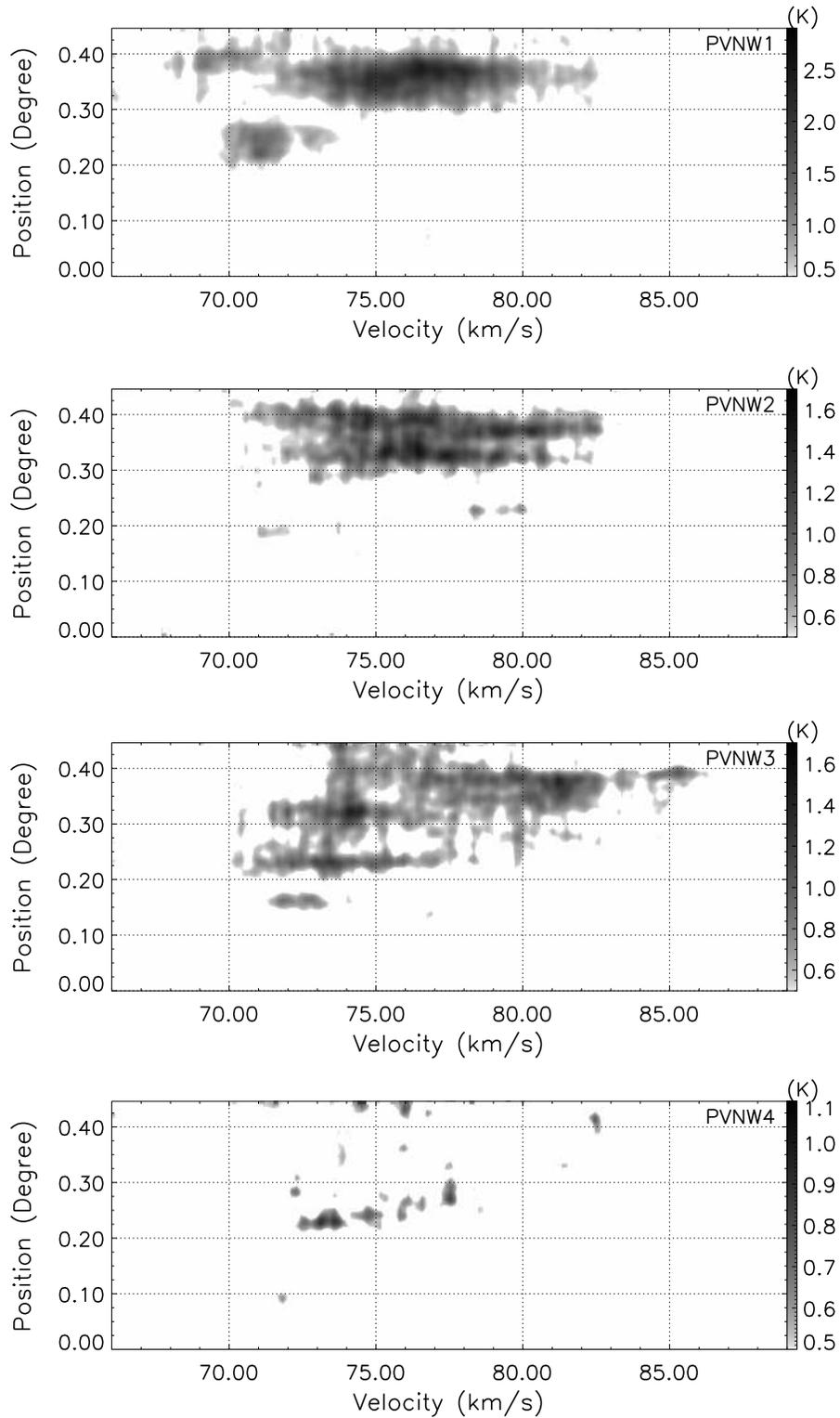}
\caption{
PV diagrams of \twCO\ ($J$=1--0) emission along PVNW1, PVNW2, PVNW3,
and PVNW4 for MC G39.315$-$1.155. The PV slices have a length of 26\farcm8 and
a width of 4\farcm5.
\label{f6}}
\end{figure}

\begin{figure}
\plotone{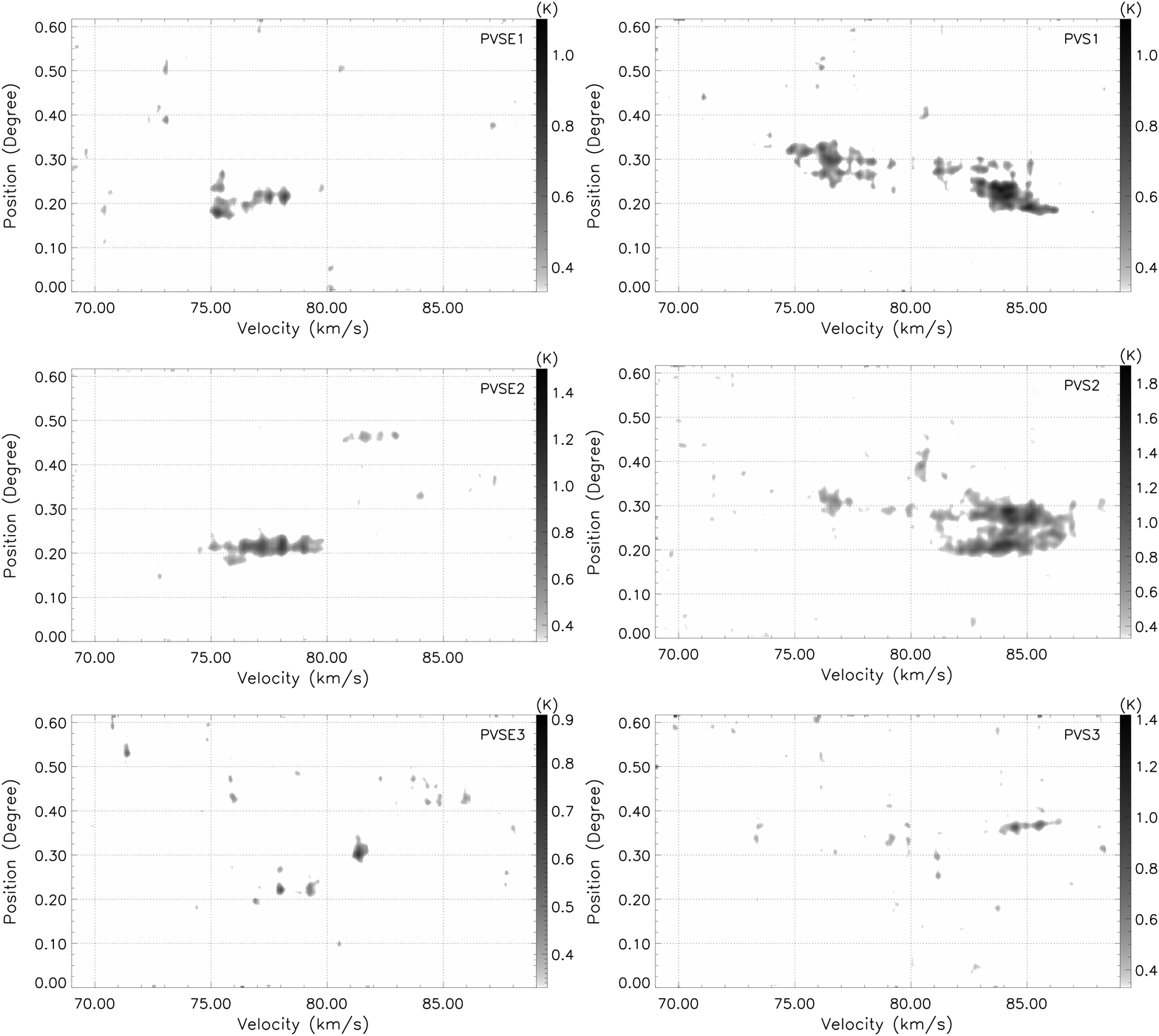}
\caption{
PV diagrams of \twCO\ ($J$=1--0) emission along PVSE1, PVSE2, PVSE3,
PVS1, PVS2, and PVS3 for MC G40.331$-$4.302. The PV slices have a length of 37\farcm0 and
a width of 4\farcm5.
\label{f7}}
\end{figure}

\begin{figure}
\plotone{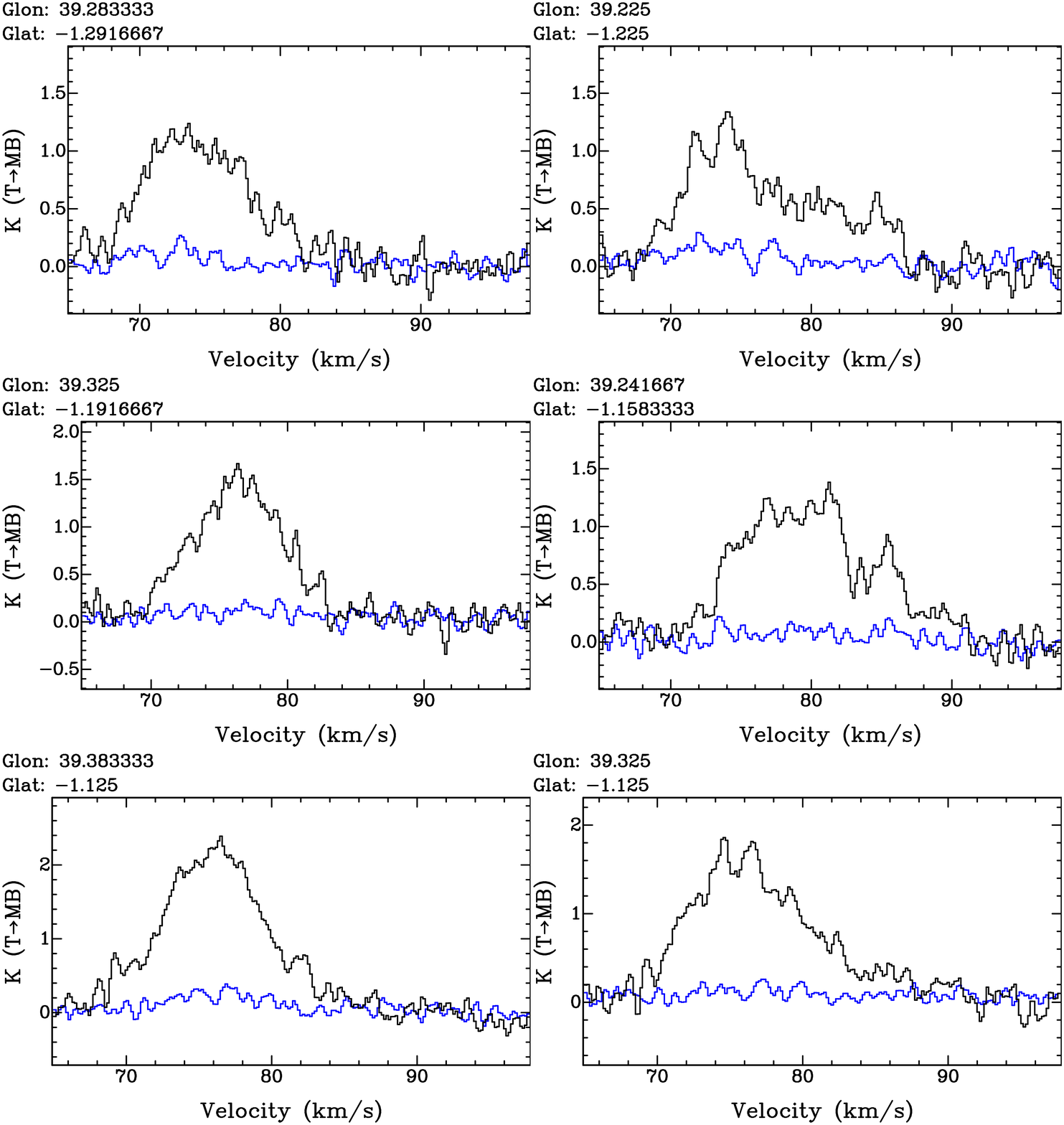}
\caption{
Typical \twCO\ ($J$=1--0; black) and \thCO\ ($J$=1--0; blue) spectra toward
MC G39.315$-$1.155.
\label{f8}}
\end{figure}

\begin{figure}
\plotone{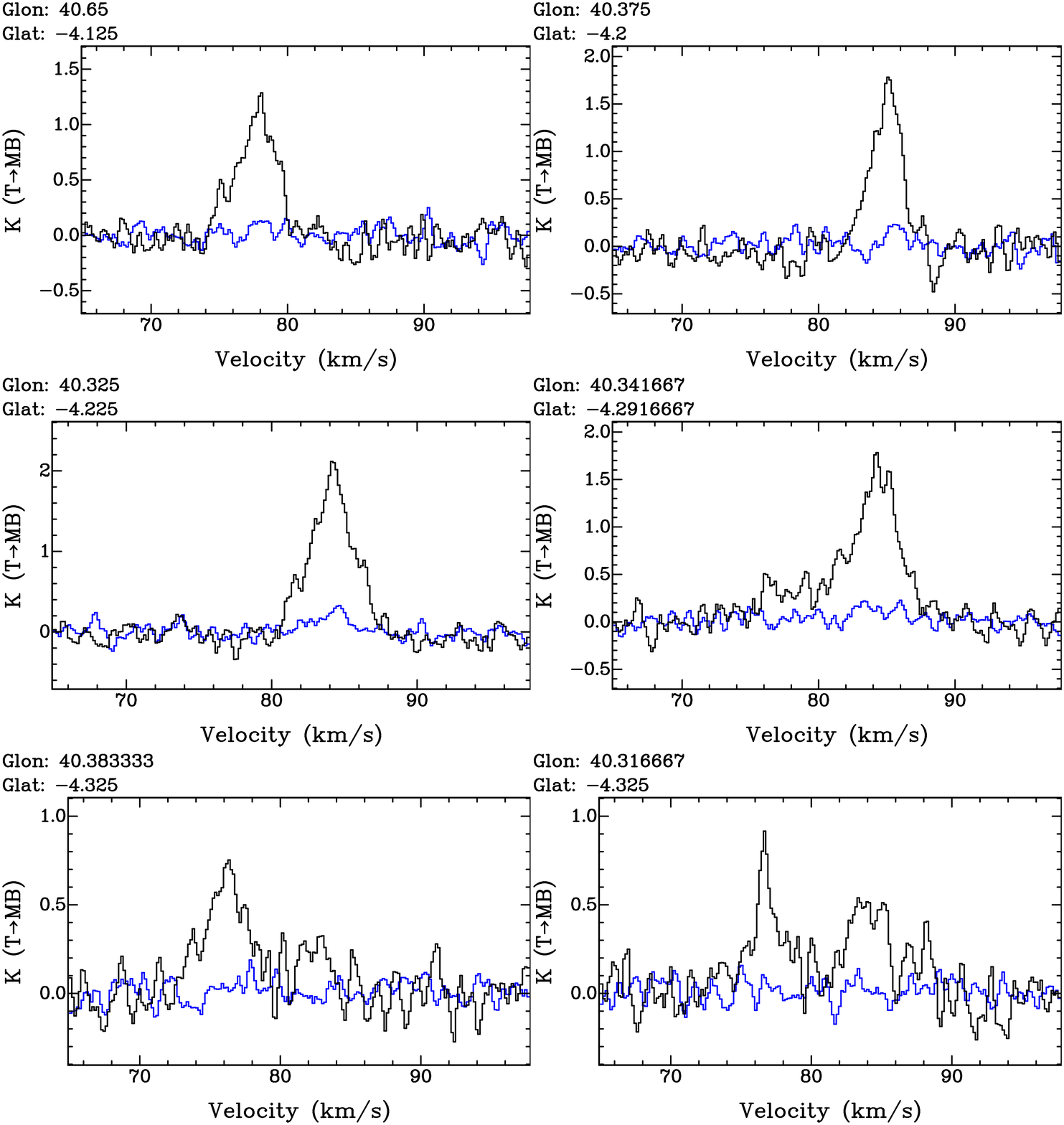}
\caption{
Typical \twCO\ ($J$=1--0; black) and \thCO\ ($J$=1--0; blue) spectra toward
MC G40.331$-$4.302.
\label{f9}}
\end{figure}

\begin{figure}
\gridline{\fig{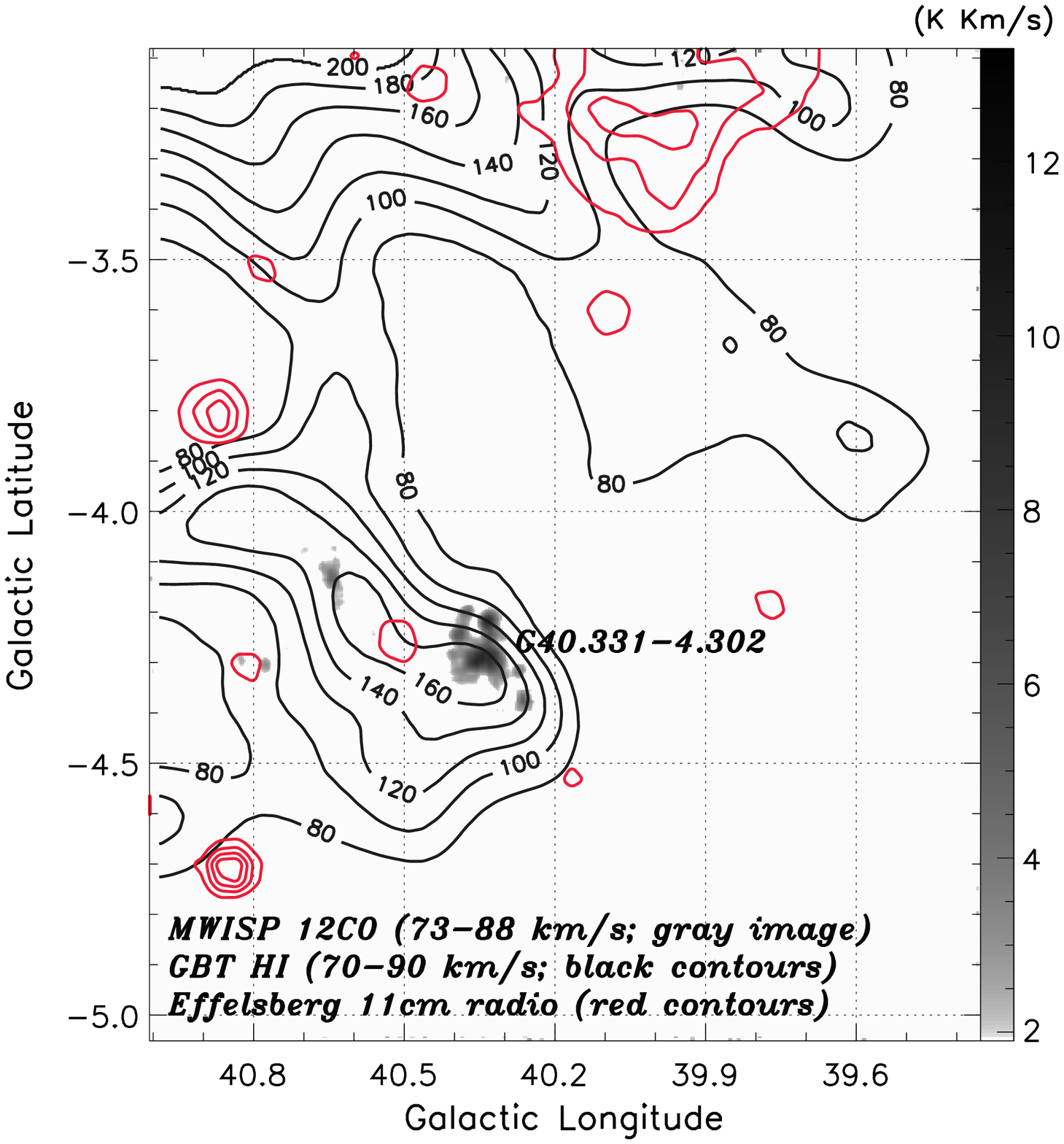}{0.55\textwidth}{}
          \fig{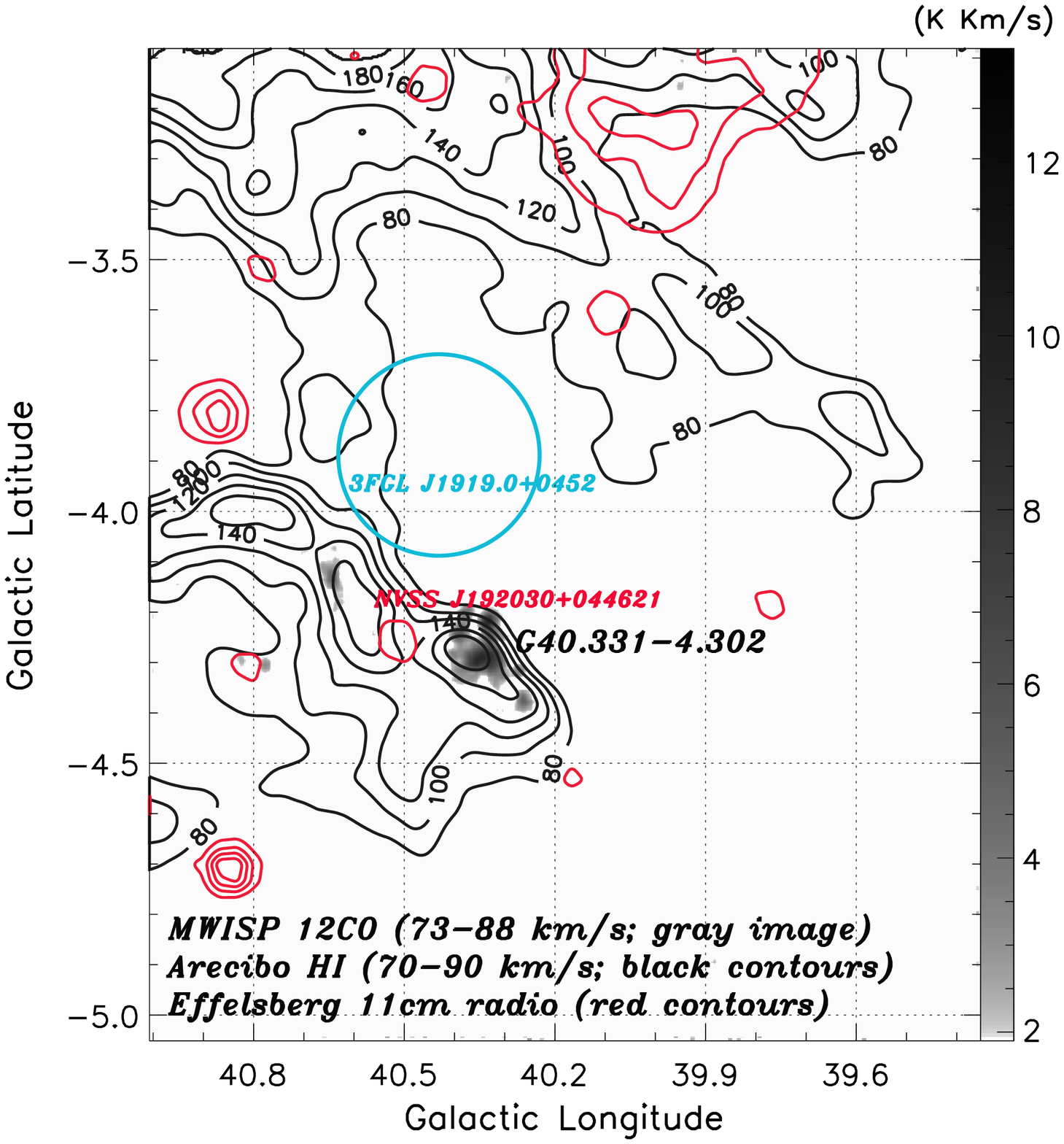}{0.55\textwidth}{}
          }
\caption{
Left panel: Integrated \twCO\ ($J$=1--0) emission 
in the interval of 73--88~km~s$^{-1}$ toward 
MC G40.331$-$4.302, overlaid with the same red radio contours as in
Figure \ref{f1}. The black contours show the GBT
\mbox{H\,\textsc{i}} emission integrated in the interval of 
70--90~km~s$^{-1}$.
Right panel: Same as the left panel but overlaid with the 
\mbox{H\,\textsc{i}} contours from the Arecibo telescope.
The cyan circle indicates the Fermi source of 3FGL J1919.0$+$0452
\citep{2015ApJS..218...23A}. The radio source of
NVSS J192030$+$044621 \citep{2010A&A...511A..53V}
is also labeled.
\label{f10}}
\end{figure}

We turn our focus on the unusual MCs G39.315$-$1.155
and G40.331$-$4.302, which are outside the current radio boundaries of W50.
Briefly, the intensity of \twCO\ emission of the two MCs is about 1--3~K and
shows non-Gaussian profiles, while the \thCO\ emission
of them is only marginally detected in some regions
with relatively strong \twCO\ emission (Figures \ref{f8}  and \ref{f9}).
Both of the two MC concentrations exhibit fragmented structures
with weak diffuse CO emission around the relatively strong CO peaks 
(see Figure \ref{f5}).
These fragmented clouds, which have somewhat different peak LSR velocities,
display broad CO-line profiles (e.g., Figures \ref{f6}--\ref{f9}).

We suggest that the two clouds are the fossil record of
interactions between the SS~433 jets and the surrounding ISM. 
The evidence are summarized as follows:

(1) Spatial coincidence. 
The two MC concentrations are exactly aligned along the precession axis of SS~433
within a small opening angle of $\sim \pm7^{\circ}$ (Section 3.4).
Both of the MCs, which appear to be elongated
from the northeast to the southwest, seem to form 
concave appearance or arc-shaped structure toward
the direction of SS~433 
(e.g., Figures \ref{f4}, \ref{f5}, and \ref{f12}).
The curvature of the two arc-like structures also points in the direction of 
SS~433. These features are similar to the
laboratory experiments \citep[e.g., see Figure 4 in][]{2002ApJ...564..113L},
magnetohydrodynamical simulations \citep[e.g.,][]{2014ApJ...789...79A},
and other astrophysical systems with jet--ISM interactions 
(\citealp[e.g., examples for the Herbig-Haro 
complex HH 1-2,][]{1998AJ....116..372H}; 
\citealp[the black hole X-ray binary GRS 1915+105,][]{2018MNRAS.475..448T};
\citealp[and the Seyfert 2 galaxy IC~5063,][]{2000AJ....119.2085O,2014Natur.511..440T,
2015A&A...580A...1M}).

(2) Kinematic features. 
The kinematics of the two clouds is consistent well with the jet properties 
of SS~433. The receding jet of SS~433 may be responsible for the 
redshifted feature of G39.315$-$1.155 and the approaching jet is
responsible for the blueshifted feature of G40.331$-$4.302 (Section 4).
The CO emission of the MCs
also exhibits broad line profiles and multi-peaks with slightly
different LSR velocities (Figures \ref{f6}--\ref{f9}). 
The angular distance of MC G39.315$-$1.155 from SS~433
is much nearer than MC G40.331$-$4.302, which agrees with 
the fact that MC G39.315$-$1.155 is becoming much more turbulent
(Section 3.4)
in the relatively high-density gas environment close to the Galactic plane.
The high turbulence and multiple gas components in the clouds 
may originate from the shock process of SS~433.

The dynamical timescale of the jet process can be estimated as 
$\sim \frac{\rm length}{\rm velocity} \gsim 2\times10^{3}$~years
when we consider the possible deceleration of the jets.
Here, the angular distance of G40.331$-$4.302 from SS~433 
is measured to be
$\sim$2\fdg15 (or $\sim$180 pc at a distance of 4.9~kpc) and the jet
velocity is assumed to be at a constant of $\lsim$0.26$c$.

On the other hand, the powerful jets of SS~433 may accumulate 
considerable material at the end of the jet shock.
For G40.331$-$4.302, the velocity difference along the LOS
(or the radial velocity component) is about
7~$\km\ps$ (Figures \ref{f7}, \ref{f9}, and \ref{f12}).
Assuming that the gas moves roughly along the precession axis of the SS 433 jets, 
the total velocity of the gas is
$\frac{\Delta V_{\rm LSR}}{{\rm cos}(i)}\sim$40~$\km\ps$,
where $i$ represents the inclination angle of the shock
to our LOS \citep[e.g., $i \approx 80^{\circ}$,][]
{1984ARA&A..22..507M,1993A&A...270..177V,2001ApJ...561.1027E}.

The atomic gas surrounding MC G40.331$-$4.302 displays multilayers 
with several $\km\ps$ velocity difference (e.g., the red, green, and blue emission
in Figure \ref{f12}).
The separation of the atomic-gas layers is measured to be $\sim 4'$ 
(or 5.7~pc at a distance of 4.9~kpc), which is approximately half 
of the thickness of the whole gas slab (Section 3.4). 
Using the above 
velocity of the shocked molecular gas of $\sim$40~$\km\ps$,
we obtain the dynamical age of the moving gas of
$\frac{3}{5} \times \frac{5.7~{\rm pc}}{40~\km\ps} \sim 0.8\times 10^{5}$~years
\citep[e.g., $\Delta l_{\rm gas}\propto t^{3/5}$,][]{1997MNRAS.286..215K}.
The dynamical age of the approaching atomic gas 
is much larger than the timescale of the current jet process,
supporting our hypothesis of previous dynamical interactions
between the SS~433 jets and its surrounding ISM at $\sim 10^{5}$~years ago.

The kinetic energy of the disturbed gas around G40.331$-$4.302 is
estimated to be $\sim 5\times 10^{49}$~erg (e.g., gas mass $\sim 3\times 10^{3} \Msun$
and velocity $\sim 40\km\ps$).
Adopting the kinetic luminosity of $\gsim 10^{39}$~erg~s$^{-1}$
for the jets \citep[e.g.,][]{2004ASPRv..12....1F,2006MNRAS.370..399B},
the kinetic energy of SS~433 are sufficient to power the
disturbed gas on the timescale of $\lsim 0.8\times10^{5}$~years for the energy
transfer efficiency of 2\%. 
On consideration of possible intermittent jet activities of SS~433, 
higher values of the kinetic energy input and/or the energy transfer 
efficiency are possible.

Recently, \cite{2017MNRAS.471.4256V} proposed that the SS~433 
system, which consists of a compact object of 
4.3$\pm 0.8 \Msun$ and a supergiant donor star of 12.3$\pm 3.3 \Msun$
\citep{2008ApJ...676L..37H}, can stably survive for the
timescale of $10^{4}-10^{5}$~years.
Because the massive donor has a radiative envelope, the system
can avoid going into a common-envelope (CE) phase and
is able to gently spiral in with stable Roche lobe overflow
\citep{1999ApJ...519L.169K,2000ApJ...530L..25K,2017MNRAS.471.4256V}. 
The timescale of $\sim 10^{4}-10^{5}$~years is controlled 
by the thermal timescale of the envelope of the 12.3 $\Msun$ 
A-supergiant in the SS~433 system \citep{2017MNRAS.471.4256V}.
We find that such the timescale is also consistent with
the average time spend in the appearance of the binary systems 
as ultraluminous X-ray sources (ULXs)
state of $\sim 10^{5}$~years \citep{2017MNRAS.465.2092P}.

For the moving atomic gas surrounding G40.331$-$4.302, 
the dynamical age of $\sim 0.8\times 10^{5}$~years estimated above is  
comparable to the evolutionary timescale of
the unusual Be/X-ray binary with stable Roche lobe overflow
\citep[e.g., for the case of the mass ratio $<$3.5,][]{2017MNRAS.471.4256V},
which is also consistent with the rapid H$_2$ formation 
timescale of $\lsim$0.1~Myr for the gas far away from the Galactic plane at 
the high pressure (see Section 6).

The previous jet episodes \citep[or possible intermittent jet activities, e.g.]
[]{2011MNRAS.414.2838G,2011MNRAS.414.2828G}, the winds from the system 
\citep[e.g.,][]{1980ApJ...238..722B,1983MNRAS.205..471K,2004ASPRv..12....1F,2006MNRAS.370..399B,
2017A&A...599A..77P},
and the loop magnetic field on the large scale \citep{2017MNRAS.467.4777F}
may play significant roles in the formation of the 
SS~433$/$W50's current configuration.

\section{Formation of molecular gas due to jet--ISM interactions?}
Figures \ref{f10}--\ref{f12} display that
a wall structure is clearly seen toward the locus of the extension of 
the precession axis of the SS~433 jets,
in which the column density of \mbox{H\,\textsc{i}} increased precipitously.
Enhanced CO emissions are also found to be nicely associated 
with the wall structure seen in \mbox{H\,\textsc{i}} emission.
MC G40.331$-$4.302 has the \mbox{H\,\textsc{i}} counterpart, which
displays more extended arc-like structure nearly perpendicular 
to the precession axis of the SS 433 jets (Figure \ref{f12}).
The positional coincidence between 
G40.331$-$4.302 and the precession axis of SS~433, as well as their coincident
kinematic features, strongly suggests the physical connection
between them (see Sections 4 and 5).

We obtain that the column density, $N_{\rm H}=N$(\mbox{H\,\textsc{i}})+$2\times N({\rm H_2})$,
is about $2.1 \times 10^{21} {\rm cm}^{-2}$ for the cloud G40.331$-$4.302,
which corresponds to $A_V\approx$1.1~mag assuming
$N_{\rm H}\approx 1.9 \times 10^{21} {\rm cm}^{-2} {\rm mag}^{-1} \times A_V$.
At the column density, the molecules are shielded from UV radiation and 
self-shielding allows significant H$_2$ molecules to exist.
The total mass of the main concentration of the cloud 
is $\sim3 \times 10^3 M_{\odot}$ within a radius of $\sim$6\farcm6 (Table 1).
The molecular fraction $f_{\rm H_2}=2N({\rm H_2})$/$N_{\rm H}$ is about 0.87,
suggesting that the dominant of the gas in the cloud is in molecular form.

The virial mass exceeds the total mass of the cloud by more than 
one order of magnitude, indicating that the cloud is in a gravitationally unbound state.
The fragmented sub-clouds (e.g., with a size of $\sim$1~pc) in MC G40.331$-$4.302 
may dissipate on a timescale of several Myr (sound crossing time) without other supports.
Alternatively, the MC may be confined by external pressure of 
$\gsim4.4 \times 10^{4}$~K~cm$^{-3}$ (e.g., $\rho (\mbox{H\,\textsc{i}}) \sigma^{2}(\mbox{H\,\textsc{i}})$)
when the magnetic pressure and the self-gravity are negligible.

We emphasize that G40.331$-$4.302 is the only detected MC within
the range of $l=$34\fdg9--45\fdg1, $-$5\fdg1 $\lsim b\lsim -$3\fdg0,
and $V_{\rm LSR}\gsim$70~km~s$^{-1}$.
It indicates that such the MC is located
about 370~pc below the Galactic plane, where a kinematic distance
of 4.9 kpc is adopted.
However, the MCs within the solar circle are concentrated in
the Galactic plane with an Gaussian FWHM of $\sim100$~pc
\citep[e.g., see Section 4.5 and Figure 6 in][]{2015ARA&A..53..583H}.
This suggests that G40.331$-$4.302 is an unusual MC
at the locus far away from the Galactic plane.
An interesting question is how to form the enormous molecular gas 
with the limited gas supply.

First of all, the formation timescale of the molecular gas, 
which was converted from the atomic gas, should be considered. 
The H$_2$ formation timescale, 
$t_{\rm H_2} \simeq \frac{10^9\ {\rm years}}{n\ ({\rm cm}^{-3})}$
\citep{1971ApJ...163..165H}, is at least 10~Myr
assuming the mean number density of $\sim$100~cm$^{-3}$ 
($n_{\rm H}+2\times n_{\rm H_2}$, see Table 1). 
However, the above estimation does not take account of 
dynamical processes, which are ubiquitous in the universe.
The processes such as shocks, turbulence, and instabilities
have a great impact on the effective H$_2$ formation rate
\citep[e.g.,][]{2007ApJ...659.1317G}.

Here, we propose that the previous jets/outflows of SS~433 may play crucial roles
in the process of molecular gas formation of G40.331$-$4.302.
The H$_2$ formation timescale can be simply written as, 
$t_{\rm H_2} \simeq 7\times10^5 f_{\rm dust} (\frac{2\times10^5\ {\rm K\ cm^{-3}}}{p_{\rm th}})^{0.95}\ 
{\rm years}$, connecting the ${\rm H_2}$ formation timescale with the thermal pressure 
\citep[here $f_{\rm dust}$ is the dust mass fraction
remaining in the gas,][]{2009A&A...502..515G}.
In our case, the ram pressure,
which is defined as $p_{\rm ram}= \rho v^2$, is probably dominant
in the clouds of interest. For the ram pressure of 
$\sim 2\times10^6\ {\rm K\ cm^{-3}}$ (e.g., $n\sim 7.3\ {\rm cm^{-3}}$ and 
$v\sim 40\ {\rm km\ s^{-1}}$), and $f_{\rm dust}$=1, the 
H$_2$ formation timescale is thus $\lsim$0.1~Myr, 
indicating very rapid MC formation. 
Several numerical models were proposed that 
${\rm H_2}$ formation occurs rapidly with dynamical processes
considered
\citep[see, e.g.,][]{2000ApJ...532..980K,2004ApJ...612..921B,
2007ApJ...659.1317G,2010MNRAS.404....2G}.
The high external pressure due to the prevailing shock 
was also suggested to accelerate the transition from atomic gas to molecular gas 
in a short time \citep[e.g.,][]{2014A&A...564A..71R,2017PASJ...69...66K}.

Furthermore, in the vicinity of G40.331$-$4.302 there may be 
considerable dark molecular gas
(DMG) \citep[e.g.,][]{2005Sci...307.1292G,2010ApJ...710..133A,
2010A&A...521L..17L,2014A&A...561A.122L,2011A&A...536A..19P},
which is nearly invisible using CO rotational emission but
considerable H$_2$ molecules may exist in CO-dark or CO-faint regions. 
Simply speaking, the scale height ($l_{\rm height}$) of the gas layer 
probably follows the relationship of 
$l_{\rm height}$(\mbox{H\,\textsc{i}})$>$$l_{\rm height}$(DMG)$>$$l_{\rm height}$(CO)  
when we see the edge-on Milky Way.
Once there exists sufficient H$_2$ molecules, CO formation may occur 
rapidly in the turbulent DMG by considering the accumulation and compression 
of the material by ram pressure from the jet.
The possible scenario can explain the unusual scale height of MC 
G40.331$-$4.302, which is nearly 370~pc below the Galactic plane.
Observations such as OH 18~cm lines, 
CH 3.3~GHz lines, and 158~$\mu$m [CII] line toward the direction 
are advocated to investigate the nature of the gas for further study.

\section{SUMMARY}
In combination with CO data from the MWISP project and \mbox{H\,\textsc{i}} data
from GBT 100~m and Arecibo 305~m telescopes,
we investigate the large-scale ISM environment toward the SS~433$/$W50 system.
Our main findings and comments are summarized as follows:

1. Two MC concentrations, MC G39.315$-$1.155 at 
$V_{\rm LSR}\sim$73--77~km~s$^{-1}$
and MC G40.331$-$4.302 at $V_{\rm LSR}\sim$77--84~km~s$^{-1}$,
are found to be well aligned along the 
precession cone of SS~433 jets within a 
smaller opening angle of $\sim \pm7^{\circ}$, 
indicating the possible connection between them. 
Further analysis show that kinematic features of the 
molecular gas are consistent with the characteristics of the SS~433
jets. That is, MC G39.315$-$1.155 toward the Galactic plane 
displays redshifted profiles for the receding gas, 
while MC G40.331$-$4.302 away from the Galactic plane 
does exhibit somewhat blueshifted features for the approaching material.

\begin{figure}
\gridline{\fig{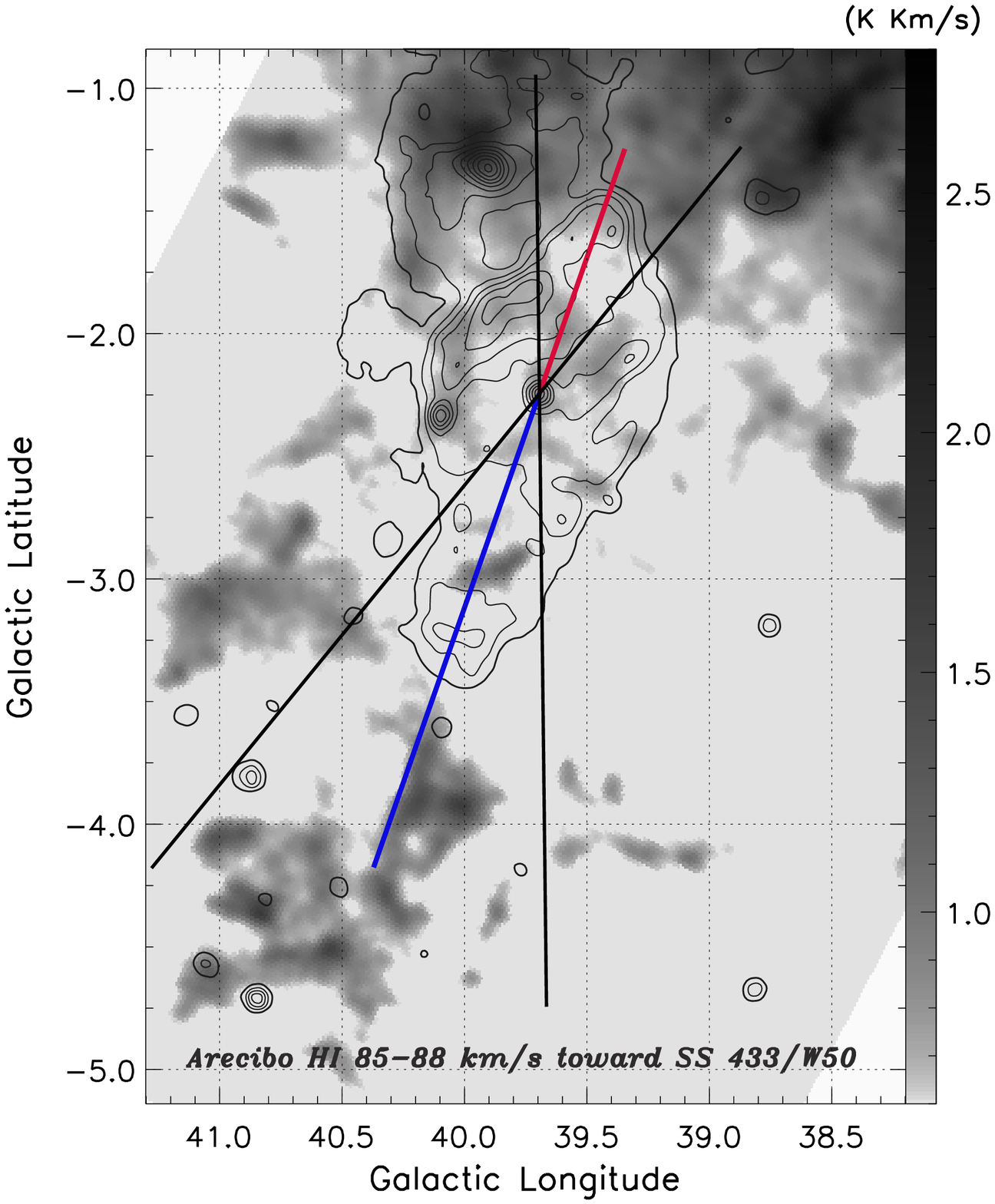}{0.55\textwidth}{(a)}
          \fig{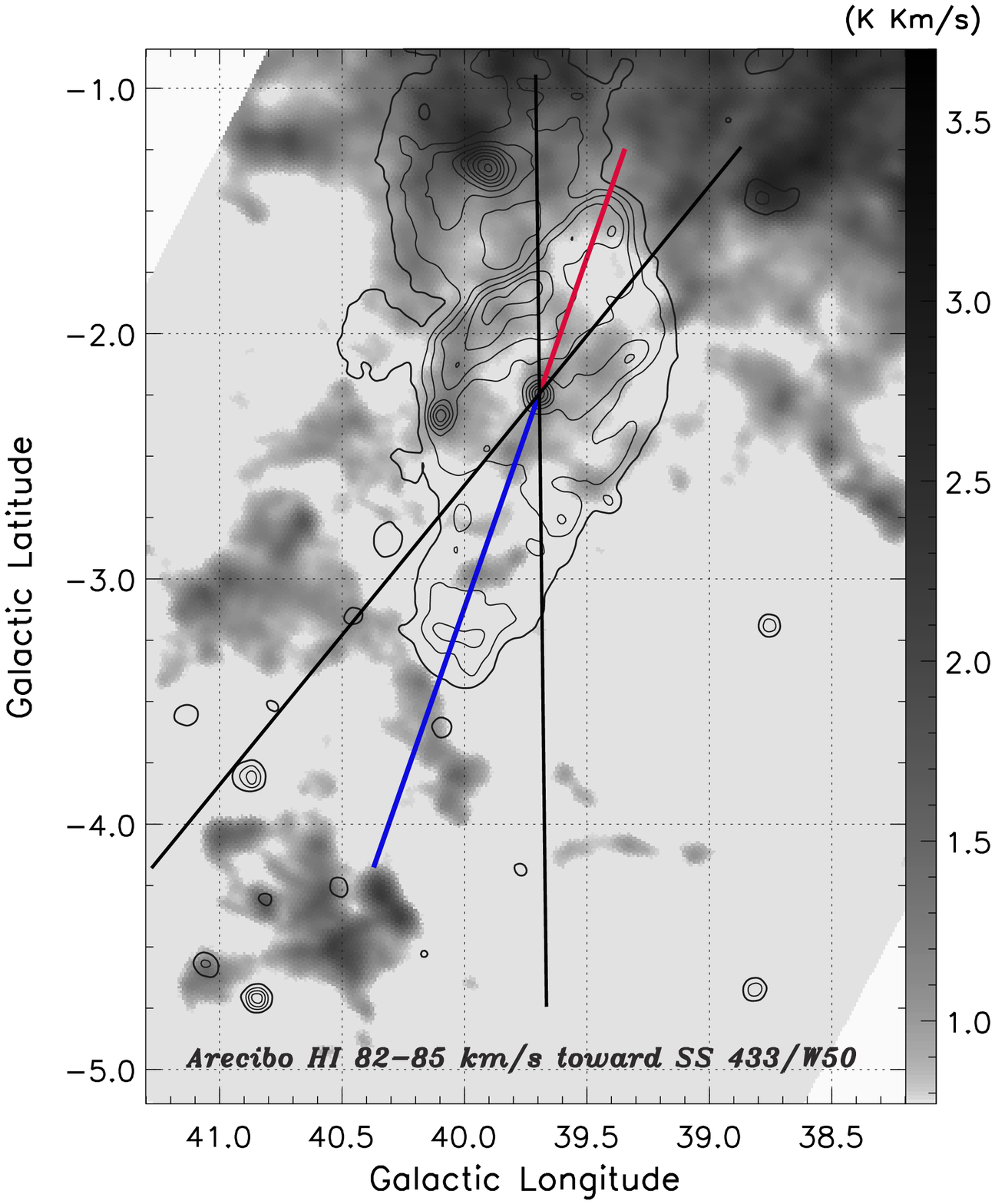}{0.55\textwidth}{(b)}
          }
\gridline{\fig{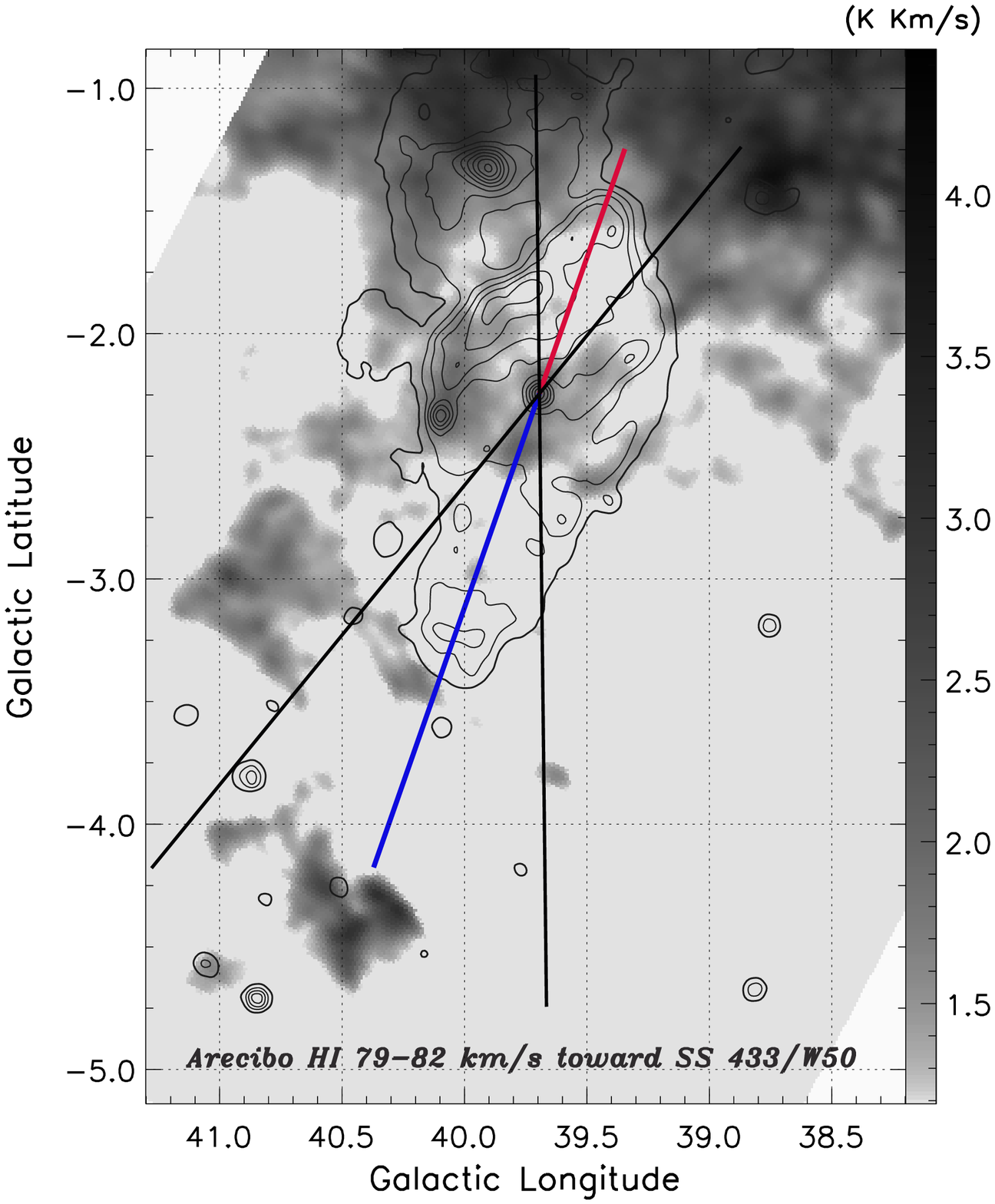}{0.55\textwidth}{(c)}
          \fig{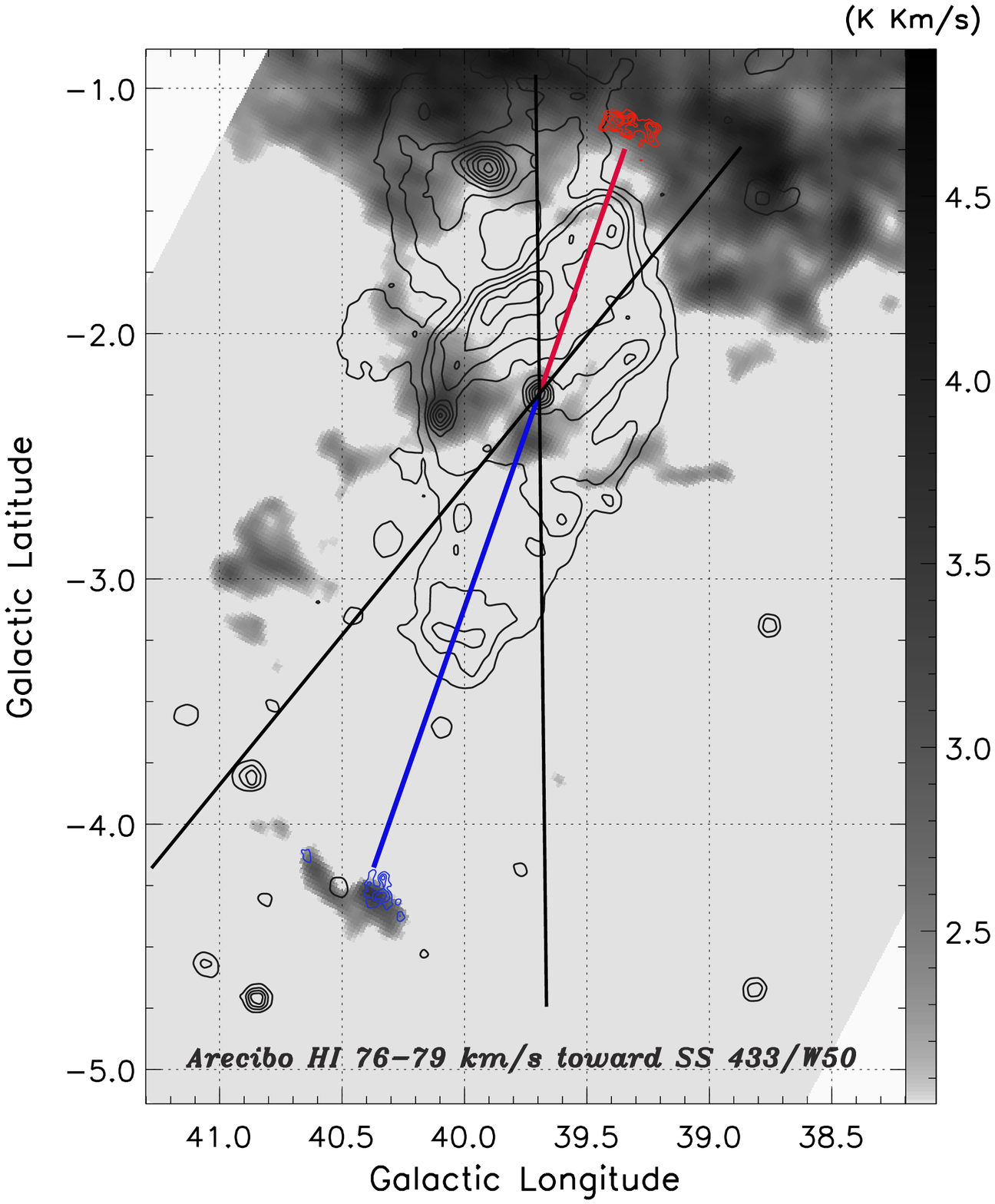}{0.55\textwidth}{(d)}
          }
\caption{
Arecibo \mbox{H\,\textsc{i}} emission integrated in
the interval of 85--88~km~s$^{-1}$, 82--85~km~s$^{-1}$,
79--82~km~s$^{-1}$, and 76--79~km~s$^{-1}$, respectively.
All images have been scaled by sin~$|b|$ to reduce the 
\mbox{H\,\textsc{i}} emission near the Galactic plane 
on a large scale and enhance the
features far from the Galactic plane.
The red and blue line indicates the precession axis of the SS 433 jets.
The overlaid black radio contours are the same as in 
Figure \ref{f1}. The red and blue contours are CO 
emission from G39.315$-$1.155 (7.5, 10.0, 12.5, 15.0, and 17.5~K~km~s$^{-1}$)
and G40.331$-$4.302 (2.5, 5.0, 7.5, and 10.0~K~km~s$^{-1}$), respectively.
\label{f11}}
\end{figure}

\begin{figure}
\plotone{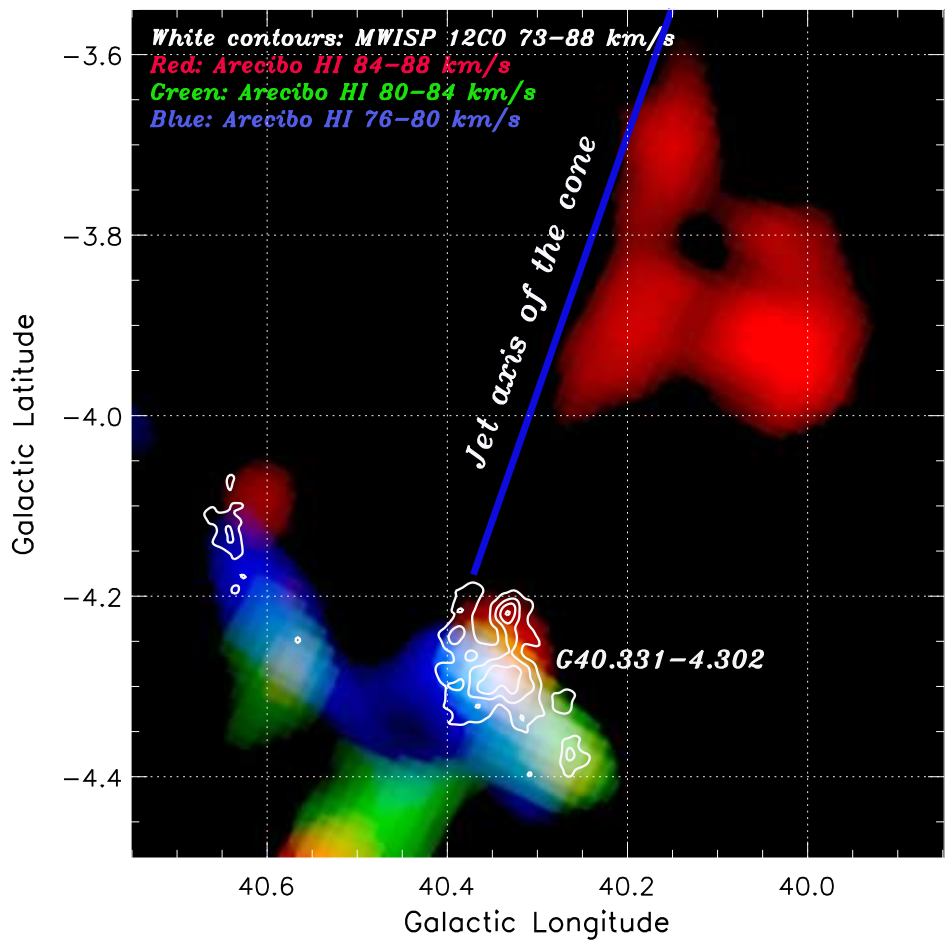}
\caption{
Arecibo \mbox{H\,\textsc{i}} emission 
(red: 84--88~km~s$^{-1}$, green: 80--84~km~s$^{-1}$, and
blue: 76--80~km~s$^{-1}$) toward MC G40.331$-$4.302.
The thick blue line indicates the precession axis of the approaching cone
of the SS 433 jets.
The CO contour levels start from 2.5 K~km~s$^{-1}$ and increase 
by a step of 2.5 K~km~s$^{-1}$.
\label{f12}}
\end{figure}

\begin{figure}
\plotone{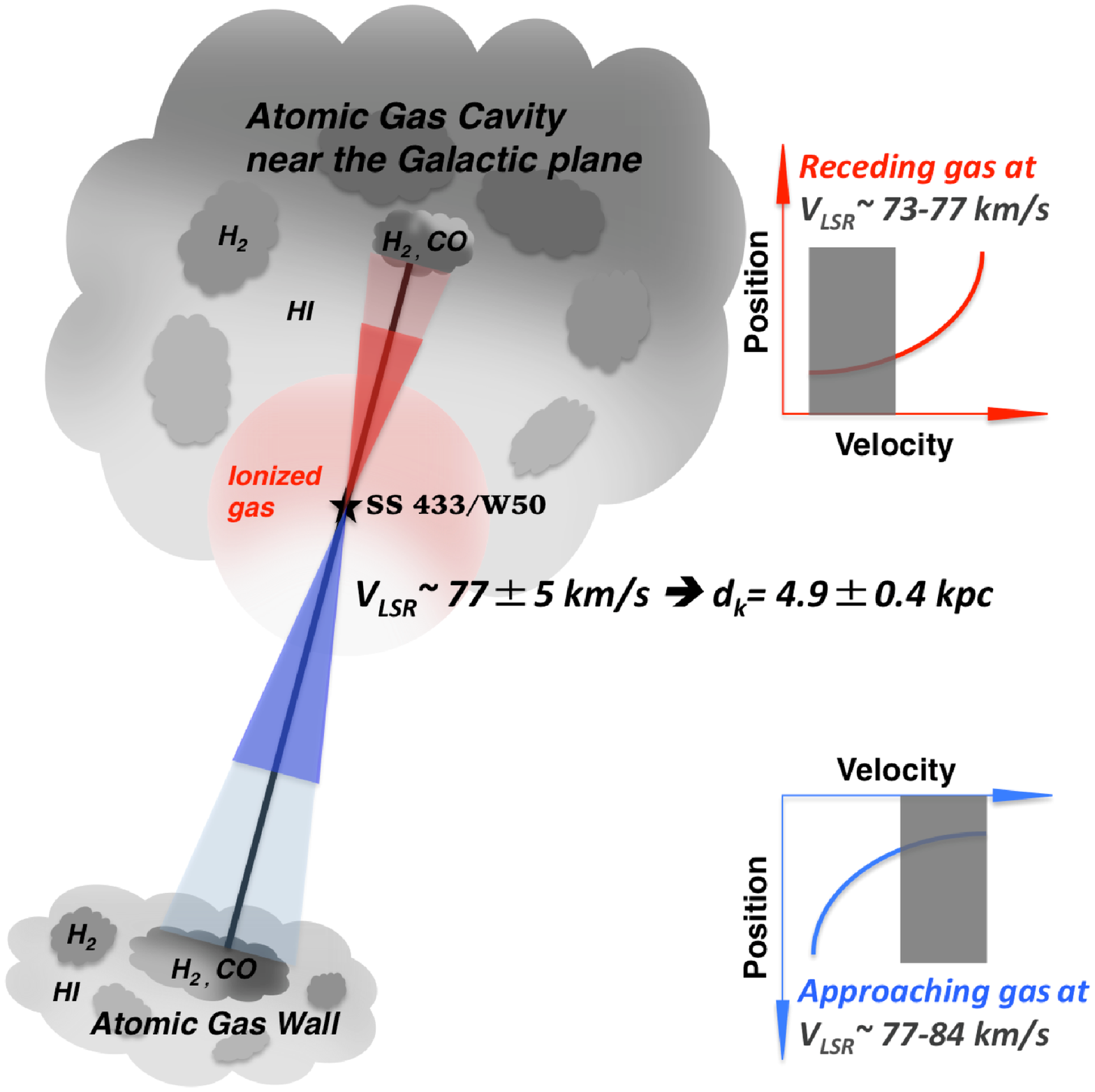}
\caption{ 
A schematic diagram toward SS~433$/$W50. 
The opening angle of the cone is about $\pm10^{\circ}$ (see Section 3.4)
around the precession axis of the SS 433 jets (thick black line).
For the right PV diagrams, the shadows represent the LSR velocity of the 
surrounding gas, while
the curves exhibit the velocity changes of the perturbed gas (see Figures \ref{f6}--\ref{f7}).
\label{f13}}
\end{figure}

2. The two MCs have corresponding atomic features traced by 
\mbox{H\,\textsc{i}} emission at $V_{\rm LSR}\sim$70--90~km~s$^{-1}$. 
For G39.315$-$1.155, the molecular gas
is located at the top of the atomic gas cavity (see panel d in Figure \ref{f11}),
although the strong \mbox{H\,\textsc{i}} emission near the Galactic 
plane prevents us from analyzing its kinematic feature of the
associated atomic gas. 
For G40.331$-$4.302, the \mbox{H\,\textsc{i}} emission also 
exhibits the blueshifted velocity structure, as well as that of
the embedded molecular gas in the apex of the atomic gas wall.
Additionally, the \mbox{H\,\textsc{i}} emission at such velocities
actually displays the morphological resemblance with the radio 
nebula W50. 

3. Based on the CO and \mbox{H\,\textsc{i}} data, the gas 
at $V_{\rm LSR}$=77$\pm$5~km~s$^{-1}$ is suggested to be physically 
associated with the SS~433$/$W50 system. 
The LSR velocity of the associated gas from radio observations
is consistent with those from optical observations.
Accordingly, the near kinematic distance of the system is
4.9$\pm$0.4~kpc, which agrees well with the results from  
the kinematic model of the proper motions of the SS 433 jets
within the error \citep[e.g., see views in][]{2014A&A...562A.130P}.
The far distance of the system can be excluded due to lacking
the \mbox{H\,\textsc{i}} absorption near
the tangent point, e.g., $V_{\rm LSR} \sim$85~km~s$^{-1}$.

4. The interesting MCs, which are exactly at the locus of the 
extension of the precession axis of the SS~433 jets, are outside the current boundaries of the 
extended nonthermal radio emission of W50. We argue that such gas is
probably related to the very early process of the SS~433 jets
and is the fossil record of jet--ISM interactions.
Analyzing the moving atomic gas surrounding MC G40.331$-$4.302,
we suggest that the dynamical process of the jet--ISM
collisions probably began at $\sim 0.8\times 10^{5}$~years ago.

5. Moreover, G40.331$-$4.302 is the only MC detected in CO emission
at such a low latitude ($\sim$370~pc away from the Galactic plane).
Connecting the blueshifted gas of G40.331$-$4.302 with the 
approaching jet of SS~433, we argue that the energetic jets of 
the microquasar have more profound effects on its
ISM environment, leading to rapid MCs formation of $\lsim$0.1~Myr
for the high ram pressure of $\sim 2\times10^6\ {\rm K\ cm^{-3}}$.
Such the rapid H$_2$ formation timescale is also 
comparable to the dynamical age of the moving atomic gas
around MC G40.331$-$4.302 
and the evolutionary timescale of the unusual binary system
\citep[e.g., a low-mass black hole and a relatively moderate-mass donor star,]
[]{2017MNRAS.471.4256V}.

6. There may be considerable H$_2$ gas with a little CO emission
in the surrounding of the fragmented MC G40.331$-$4.302.
We are advocating that further observations such as OH 18~cm lines,
CH 3.3~GHz lines, and 158~$\mu$m [CII] line are helpful 
in investigating the nature of the CO-faint MC 
and the nearby regions.

\acknowledgments
The authors acknowledge the staff members of the Qinghai Radio
Observing Station at Delingha for their support of the observations.
We would like to thank the anonymous referee for valuable comments
and suggestions that helped to improve this paper.
This work is supported by the National Key R\&D Program of China
through grants 2017YFA0402701, 2017YFA0402702, 2017YFA0402600 and 2015CB857100.
Y.S. thanks F. J. Lockman for providing the GBT HI data
in our preliminary study. Y.C. thanks the National Natural Science Foundation of China
for the support through grants 11773014, 11633007 and 11851305.
Y.C. acknowledges support by the NSFC through grant 11473069.
This publication utilizes data from Galactic ALFA HI (GALFA HI) survey 
data set obtained with the Arecibo L-band Feed Array (ALFA) on the 
Arecibo 305m telescope. Arecibo Observatory is part of the National 
Astronomy and Ionosphere Center, which is operated by Cornell 
University under Cooperative Agreement with the U.S. National Science 
Foundation. The GALFA HI surveys are funded by the NSF through grants 
to Columbia University, the University of Wisconsin, and the 
University of California.

\facility{PMO 13.7m}
\software{GILDAS/CLASS \citep{2005sf2a.conf..721P}} 

\bibliographystyle{aasjournal}
\bibliography{references}

\begin{deluxetable*}{cccccccc}[b!]
\tablecaption{Properties of Gas toward G39.315$-$1.155 and G40.331$-$4.302}
\tablecolumns{8}
\tablenum{1}
\tablewidth{0pt}
\tablehead{
\colhead{Name} &
\colhead{Tracer} &
\colhead{Area\tablenotemark{a}} &
\colhead{Radius\tablenotemark{b}} &
\colhead{$I_{\rm mean}$\tablenotemark{c}} &
\colhead{Column Density\tablenotemark{d}} &
\colhead{Mass\tablenotemark{e}}    &
\colhead{Volume Density\tablenotemark{e,f}} \\
\colhead{}              &
\colhead{}              & \colhead{(arcmin$^2$)} & \colhead{(arcmin)} &
\colhead{(K~$\km\ps$)}  & \colhead{($\times10^{20}$~cm$^{-2}$)}  & 
\colhead{($\times10^{3} M_{\odot}$)} & \colhead{(cm$^{-3}$)}
}
\startdata
G39.315$-$1.155  & \twCO\                &  136.5  &  6.6  &    8.8   &  $N({\rm H_2})=17.6$ &
$M({\rm H_2})=10.7d_{4.9}^2$   &  $n({\rm H_2})=45.5d_{4.9}^{-1}$  \\  
\hline
G40.331$-$4.302  & \mbox{H\,\textsc{i}}  & 135.0  &  6.6  &  154.2   &  $N({\rm H})=2.8$   &
$M({\rm H})=0.8d_{4.9}^2$   &   $n({\rm H})=7.3d_{4.9}^{-1}$  \\      
		 & \twCO\                &  53.3  &  4.1  &    4.6   &  $N({\rm H_2})=9.2$ &
$M({\rm H_2})=2.2d_{4.9}^2$   &  $n({\rm H_2})=38.2d_{4.9}^{-1}$  \\  
\enddata
\tablenotetext{a}{The area of the emission with $I$(CO)$>$5~K~$\km\ps$ for G39.315$-$1.155
and $I$(\mbox{H\,\textsc{i}})$>$130~K~$\km\ps$ and $I$(CO)$>$2.5~K~$\km\ps$ for G40.331$-$4.302.}
\tablenotetext{b}{The effective radius is calculated from (Area/3.14)$^{0.5}$.}
\tablenotetext{c}{The mean intensity of the \mbox{H\,\textsc{i}} 
(70--90~$\km\ps$) and CO (73--88~$\km\ps$) emission.}
\tablenotetext{d}{The column density of the atomic and molecular gas is calculated from 
1.823$\times10^{18}\times I_{\rm mean}$(\mbox{H\,\textsc{i}}) and
2$\times10^{20}\times I_{\rm mean}$(CO), respectively. See text.}
\tablenotetext{e}{Parameter $d_{4.9}$ is the distance to the cloud
in units of 4.9~kpc (see Section 4).}
\tablenotetext{f}{The volume density of the atomic and molecular gas is calculated from
the mass and the effective radius of the emission.}
\end{deluxetable*}

\end{document}